\documentclass[twoside,11pt]{article}

\usepackage{jmlr2e}

\pdfoutput=1
\usepackage{url}            
\usepackage{booktabs}       
\usepackage{nicefrac}       
\usepackage{microtype}      
\usepackage[font=small,skip=0pt,labelfont=bf]{caption}
\usepackage{mdwmath, mdwtab}
\usepackage{amsmath, amssymb, amsbsy}
\usepackage[caption=false,font=footnotesize]{subfig}
\usepackage[linesnumbered,ruled,vlined]{algorithm2e}
\usepackage{array}
\usepackage{eqparbox}
\usepackage{filecontents}
\usepackage{comment}
\usepackage{lipsum}
\usepackage{xcolor}

\newtheorem{assumption}{Assumption}

\newcommand{\matrx}[1]{\ensuremath\boldsymbol{\rm #1}}
\newcommand{\vect}[1]{\ensuremath\boldsymbol{\rm #1}}

\newcommand{\twonorm}[1]{\ensuremath{\|#1\|_2}}

\newcommand{\ceil}[1]{\lceil#1\rceil}
\newcommand{\bvec}[1]{\ensuremath{\mathbf{#1}}}
\newcommand{\gvec}[1]{\ensuremath{\mbox{\boldmath$#1$}}} 
\newcommand{\norm}[1]{\left\lVert#1\right\rVert}
\DeclareMathOperator*{\argmin}{\arg\!\min}
\DeclareMathOperator*{\argmax}{\arg\!\max}

\allowdisplaybreaks[4]

\usepackage{xspace}
\newcommand{\eg}{\emph{e.g.,}\xspace}
\newcommand{\ie}{\emph{i.e.,}\xspace}
\newcommand{\passcode}{\emph{PassCoDe}\xspace}

\newcommand{\cocoap}{\emph{CoCoA}+\xspace}
\newcommand{\baseline}{\emph{Baseline}\xspace}
\newcommand{\hdca}{\emph{Hybrid-DCA}\xspace}

\def\mytitle{Hybrid-DCA: A Double Asynchronous Approach for Stochastic Dual Coordinate Ascent}

\jmlrheading{x}{yyyy}{pp-pp}{mm/yy}{mm/yy}{Soumitra Pal, Tingyang Xu, Tianbao Yang, Sanguthevar Rajasekaran and Jinbo Bi}

\ShortHeadings{Hybrid-DCA: Double Asynchronous Stochastic DCA}{Pal, Xu, Yang, Rajasekaran, Bi}
\firstpageno{1}

\title{\mytitle}
\author{\name Soumitra Pal \email mitra@uconn.edu\\ 
  \name Tingyang Xu \email tix11001@engr.uconn.edu\\
  \addr Computer Science and Engineering,  University of Connecticut, Storrs, CT 06269 USA
  \AND
  \name Tianbao Yang \email tianbao-yang@uiowa.edu\\
  \addr Computer Science, University of Iowa, Iowa City, IW 52242 USA
  \AND
  \name Sanguthevar Rajasekaran \email rajasek@engr.uconn.edu\\
  \name Jinbo Bi \email jinbo@engr.uconn.edu\\
  \addr Computer Science and Engineering,  University of Connecticut, Storrs, CT 06269 USA
}

\editor{Not known yet}

\begin{document}

\maketitle

\begin{abstract}
  In prior works, stochastic dual coordinate ascent (SDCA) has been parallelized in a multi-core environment where the cores communicate through shared memory, or in a multi-processor distributed memory environment where the processors communicate through message passing. In this paper, we propose a hybrid SDCA framework for multi-core clusters, the most common high performance computing environment that consists of multiple nodes each having multiple cores and its own shared memory. We distribute data across nodes where each node solves a local problem in an asynchronous parallel fashion on its cores, and then the local updates are aggregated via an asynchronous across-node update scheme. The proposed double asynchronous method converges to a global solution for $L$-Lipschitz continuous loss functions, and at a linear convergence rate if a smooth convex loss function is used. Extensive empirical comparison has shown that our algorithm scales better than the best known shared-memory methods and runs faster than previous distributed-memory methods. Big datasets, such as one of 280 GB from the LIBSVM repository, cannot be accommodated on a single node and hence cannot be solved by a parallel algorithm. For such a dataset, our hybrid algorithm takes 30 seconds to achieve a duality gap of $10^{-6}$ on 16 nodes each using 8 cores, which is significantly faster than the best known distributed algorithms, such as CoCoA+, that take more than 300 seconds on 16 nodes.
\end{abstract}

\begin{keywords}
  dual coordinate descent, distributed computing, optimization
\end{keywords}

\section{Introduction}
\label{sec:intro}

The immense growth of data has made it important to efficiently solve large scale machine learning problems. It is necessary to take advantage of modern high performance computing (HPC) environments such as multi-core settings where the cores communicate through shared memory, or multi-processor distributed memory settings where the processors communicate by passing messages. In particular, a large class of supervised learning formulations, including support vector machines (SVMs), logistic regression, ridge regression and many others, solve the following generic regularized risk minimization (RRM) problem: given a set of instance-label pairs of data points $(\bvec{x}_i, y_i), i=1,\ldots, n$,
\begin{align}
  \min_{\bvec{w} \in \mathbb{R}^d} P(\bvec{w}) := \frac{1}{n}\sum_{i=1}^{n} \phi(\bvec{x}_i^{\top} \bvec{w}; y_i) + \frac{\lambda}{2} g({\bvec{w}}),
\label{eq:primal}
\end{align}
where $y_i {\in} \mathbb{R}$ is the label for the data  point $\bvec{x}_i \in \mathbb{R}^d$,
$\bvec{w} \in \mathbb{R}^d$ is the linear predictor to be optimized, $\phi$ is a loss function that is convex with respect to its first argument, $\lambda$ is a regularization parameter that balances between the loss and a regularizer $g(\bvec{w})$, especially the $\ell_2$-norm penalty $\twonorm{\bvec{w}}^2$.

Many efficient sequential algorithms have been developed in the past decades to solve (\ref{eq:primal}), \eg stochastic gradient descent (SGD)~\cite{zhang2004solving}, or alternating direction method of multipliers (ADMM)~\cite{boyd2011distributed}. Especially, (stochastic) dual coordinate ascent (DCA) algorithm~\cite{shalev2013stochastic} has been one of the most widely used algorithms for solving~\eqref{eq:primal}. It efficiently optimizes the following dual formulation (\ref{eq:dual})
\begin{align}
  \max_{\gvec{\alpha} \in \mathbb{R}^n} D(\gvec{\alpha}) :=  - \frac{1}{n}\sum_{i=1}^{n} \phi^*(-\alpha_i) -\frac{\lambda}{2} g^*\left(\frac{1}{\lambda n} \matrx X \gvec{\alpha}\right),
\label{eq:dual} \\
  \text{using}\qquad
  \bvec{w}(\gvec{\alpha}) = \nabla g^*\left(\frac{1}{\lambda n}\matrx X \gvec{\alpha}\right),
  \label{eq:relation}
\end{align}
where $\phi^*,g^*$ are the convex conjugates of $\phi, g$, respectively, defined as, \eg $\phi^*(u) = \max_z(zu - \phi(z))$ and it is known that if $\gvec{\alpha}^*$ is an optimal dual solution then $\bvec{w}^* = \bvec{w}(\gvec{\alpha}^*)$ is an optimal primal solution and $P(\bvec{w}^*) = D(\gvec{\alpha}^*)$. The dual objective has a separate dual variable associated with each training data point. The stochastic DCA updates dual variables, one at a time, while maintaining the primal variables by calculating~\eqref{eq:relation} from the dual variables.

Recently, many efforts have been undertaken to solve~\eqref{eq:primal} in a distributed or parallel framework. It has been shown that distributed DCA algorithms have comparable and sometimes even better convergence than SGD-based or ADMM-based distributed algorithms~\cite{yang2013trading}.  The distributed DCA algorithms can be grouped into two sets. The first set contains synchronous algorithms in which a random dual variable is updated by each processor and the primal variables are synchronized across the processors in every iteration~\cite{jaggi2014communication,ma2015adding,yang2013trading}. This approach incurs a large communication overhead. The second set of algorithms avoids communication overhead by exploiting the shared memory in a multi-core setting~\cite{hsieh2015passcode} where the primal variables are stored in a primary memory shared across all the processors. Further speedups have been obtained by using (asynchronous) atomic memory operations instead of costly locks for shared memory updates~\cite{hsieh2015passcode,peng2015arock}. Nevertheless, this approach is difficult to scale up for big datasets that cannot be fully accommodated in the shared memory. This leads to a challenging question: how do we scale up the asynchronous shared memory approach for big data while maintaining the speed up?   
\begin{figure}
  \centering
  \subfloat[\label{subfig-1:dummy}]{%
    \includegraphics[width=0.55\linewidth]{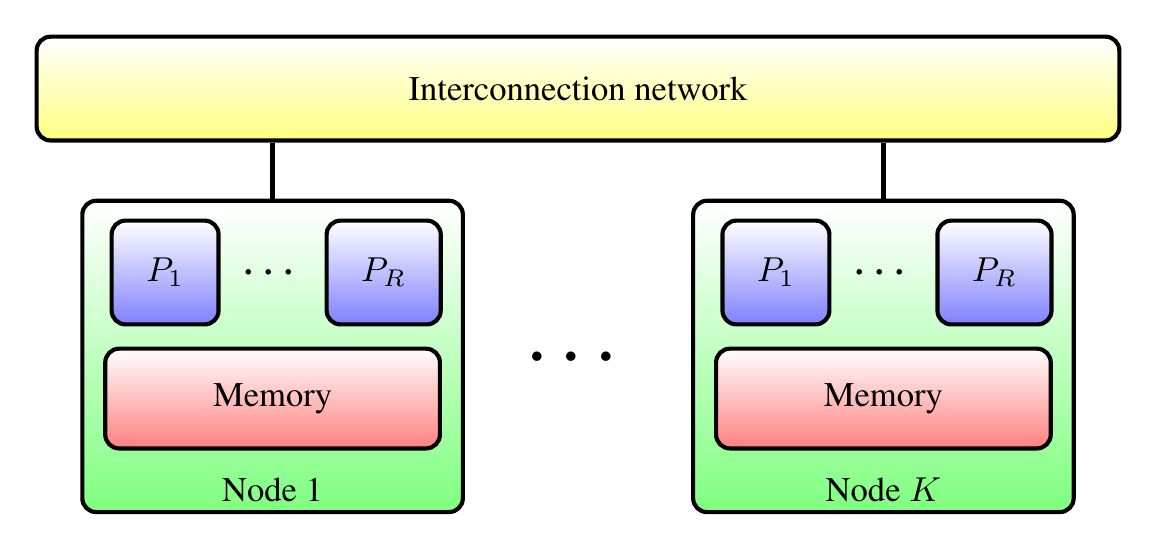}
  }
  \hfill
  \subfloat[\label{fig:spectrum}]{%
    \includegraphics[width=0.4\linewidth]{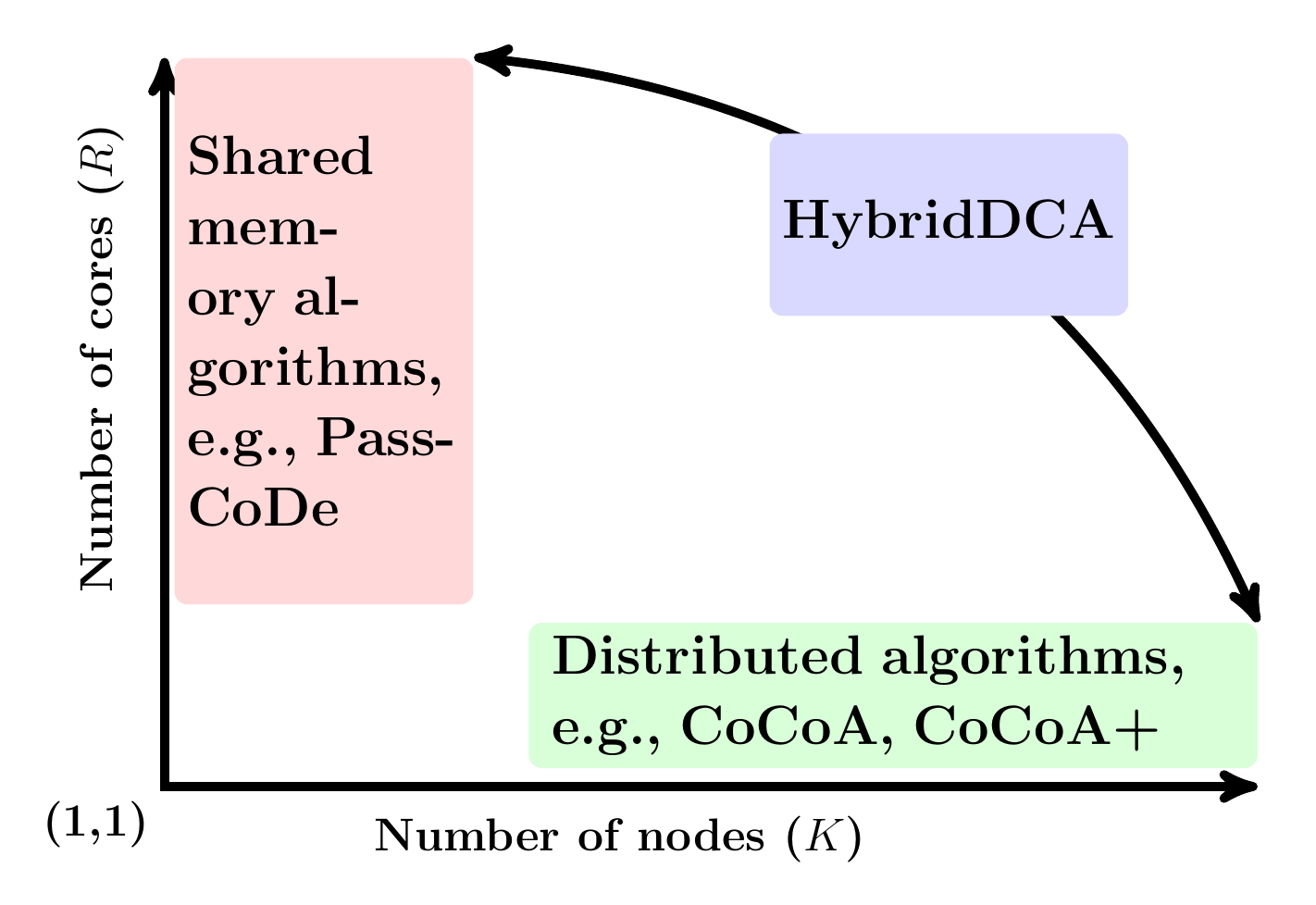}
  }
  \caption{(a) A simplified view of the modern HPC system and (b) Algorithms on this architecture.}
  \label{fig:hpcarch}
\end{figure}

We address this challenge by proposing and implementing a hybrid strategy. The modern HPC platforms can be viewed as a collection of $K$ nodes interconnected through a network as shown in Fig.~\ref{fig:hpcarch}(a). Each node contains a memory shared among $R$ processing cores. Our strategy exploits this architecture by equally distributing the data across the local shared memory of the $K$ nodes.  Each of the $R$ cores within a node runs a computing thread that asynchronously updates a random dual variable from those associated with the data allocated to the node. Each node also runs a communicating thread. One of the communicating threads is designated as \emph{master} and the rest are \emph{workers}. After every round of $H$ local iterations in each computing thread, each worker thread sends the local update to the master. After accumulating the local updates from $S$ of the $K$ workers, the master broadcasts the global update to the contributing workers. However, to avoid a slower worker falling back too far, the master ensures that in every $\Gamma$ consecutive global updates there is at least one local update from each worker.
Fig.~\ref{fig:hpcarch}(b) shows how our scheme is a generalization of the existing approaches: for $K=1$, our setup coincides with the shared memory multi-core setting~\cite{hsieh2015passcode} and for $R=1, S=K$ our setup coincides with the synchronous algorithms in distributed memory setting~\cite{jaggi2014communication,ma2015adding,yang2013trading}. With a proper adjustment of the parameters $H, S, \Gamma$ our strategy could balance the computation time of the first setting with the communication time of the second one, while ensuring scalability in big data applications.

Thus, our contributions are 1) we propose
and analyze a hybrid asynchronous shared memory and asynchronous distributed memory implementation (\emph{Hybrid-DCA}) of the mostly used DCA algorithm to solve (\ref{eq:primal});
2) we prove a strong guarantee of convergence for $L$-Lipschitz continuous loss functions, and further linear convergence when a smooth convex loss function is used; and 3) the experimental results using our light-weight OpenMP+MPI implementation show that our algorithms are much faster than existing distributed memory algorithms~\cite{jaggi2014communication,ma2015adding}, and easily scale up with the volume of data in comparison with the shared memory based algorithms~\cite{hsieh2015passcode} as the shared memory size is limited. 

\section{Related Work}
\label{sec:survey}

\textbf{Sequential Algorithms.} SGD is the oldest and simplest method of solving problem~\eqref{eq:primal}. Though SGD is easy to implement and converges to modest accuracy quickly, it requires a long tail of iterations to reach `good' solutions and also requires adjusting a step-size parameter. On the other hand, SDCA methods are free of learning-rate parameters and have faster convergence rate around the end~\cite{moulines2011non,needell2014stochastic}. A modified SGD has also been proposed with faster convergence by switching to SDCA after quickly reaching a modest solution~\cite{shalev2013stochastic}. Recently, `variance reduced' modifications to the original SGD have also caught attention. These modifications estimate gradients with small variance as they approach to an optimal solution. Mini-batch algorithms are also proposed to update several dual variables (data points) in a batch rather than a single data point per iteration\cite{takac2013mini}. Mini-batch versions of both SGD and SDCA have slower convergence when the batch size increases\cite{richtarik2013distributed,shalev2016accelerated}. All these sequential algorithms become ineffective when the datasets get bigger.



\textbf{Distributed Algorithms.} In the early single communication scheme~\cite{mcwilliams2014loco,heinze2015dual,mcdonald2009efficient}, a dataset is `decomposed' into smaller parts that can be solved independently. The final solution is reached by `accumulating' the partial solutions using a single round of communications. This method has limited utility because most datasets cannot be decomposed in such a way. 
Using the primal-dual relationship~\eqref{eq:relation}, fully distributed algorithms of DCA are later developed where each processor updates a separate $\alpha_i$ which is then used to update $\vect w(\vect \alpha)$, and synchronizes $\vect w$ across all processors (\eg CoCoA~\cite{jaggi2014communication}). To trade off communications vs computations, a processor can solve its subproblem with $H$ dual updates before synchronizing the primal variable (\eg CoCoA+~\cite{ma2015adding},DisDCA~\cite{yang2013trading}). In~\cite{yang2013trading,ma2015adding}, a more general framework is proposed in which the subproblem can be solved using not only SDCA but also any other sequential solver that can guarantee a $\Theta$-approximation of the local solution for some $\Theta \in (0,1]$. Nevertheless, the synchronized update to the primal variables has the inherent drawback that the overall algorithm runs at a speed of the slowest processor even when there are fast processors~\cite{agarwal2014reliable}.

\textbf{Parallel Algorithms.} Multi-core shared memory systems have also been exploited, where the primal variables are maintained in a shared memory, removing the communication cost. However, updates to shared memory requires synchronization primitives, such as locks, which again slows down computation. Recent methods~\cite{hsieh2015passcode,liu2015asynchronous} avoid locks by exploiting (asynchronous) atomic memory updates in modern memory systems. There is even a wild version~in~\cite{hsieh2015passcode} that takes arbitrarily one of the simultaneous updates. Though the shared memory algorithms are faster than the distributed versions, they have an inherent drawback of being not~scalable, as there can be only a few cores in a processor board. 

\textbf{Other Distributed Methods for RRM.} Besides distributed DCA methods, there are several recent distributed versions of other algorithms with faster convergence, including distributed Newton-type methods (DISCO~\cite{DBLP:journals/corr/ZhangX15a}, DANE~\cite{DBLP:conf/icml/ShamirS014}) and distributed stochastic variance reduced gradient method (DSVRG~\cite{DBLP:journals/corr/LeeLin15}). It has  been shown that they can achieve the same accurate solution using  fewer rounds of communication, however, with additional computational overhead. In particular, DISCO and DANE need to solve a linear system in each round, which could be very expensive for higher dimensions. DSVRG requires each machine to load and store a second subset of the data sampled from the original training data, which also increase its running time.

The ADMM~\cite{boyd2011distributed} and quasi-Newton methods such as L-BFGS also have distributed solutions. These methods have low communication cost, however, their inherent drawback of computing the full batch gradient does not give computation vs communications trade-off.  In the context of consensus optimization, \cite{zhang2014asynchronous} gives an asynchronous distributed ADMM algorithm but that does not directly apply to solving~\eqref{eq:primal}.

To the best of our knowledge, this paper is the first to propose, implement and analyze a hybrid approach exploiting modern HPC architecture. Our approach is the amalgamation of three different ideas -- 1) CoCoA+/DisDCA distributed framework, 2) asynchronous multi-core shared-memory solver~\cite{hsieh2015passcode} and 3) asynchronous distributed approach~\cite{zhang2014asynchronous} -- taking the best of each of them. In a sense ours is the first algorithm which asynchronously uses updates which themselves have been computed using asynchronous methods. 

\section{Algorithm}
\label{sec:algos}

At the core of our algorithm, the data is distributed across $K$ nodes and each node, called a \emph{worker}, repeatedly solves a perturbed dual formulation on its data partition and sends the local update to one of the nodes designated as the \emph{master} which merges the local updates and sends back the accumulated global update to the workers to solve the subproblem once again, unless a global convergence is reached.
Let $\mathcal{I}_k \subseteq \{1, 2, \ldots, n\}, k=1, \ldots, K$ denote the indices of the data and the dual variables residing on node $k$ and $n_k = |\mathcal{I}_k|$. For any $\vect z \in \mathbb{R}^n$ let $\vect z_{[k]}$ denote the vector in $\mathbb{R}^{n}$ defined in such a way that the $i$th component $(\vect z_{[k]})_i = z_i$ if $i \in \mathcal{I}_k$, $0$ otherwise. Let $\matrx X_{[k]} \in \mathbb{R}^{d \times n}$ denote the matrix consisting of the columns of the $\matrx X \in \mathbb{R}^{d \times n}$ indexed by $\mathcal{I}_k$ and replaced with zeros in all other columns.

Ideally, the dual problem solved by node $k$ is \eqref{eq:dual} with $\matrx X, \vect \alpha$ replaced by $\matrx X_{[k]}, \vect \alpha_{[k]}$, respectively, and hence is independent of other nodes. However, following the efficient practical implementation idea in~\cite{yang2013trading,ma2015adding}, we let the workers communicate among them a vector $\vect v \in \mathbb{R}^d$, an estimate of $\vect w(\vect \alpha) = \frac{1}{\lambda n} \matrx X \vect \alpha$ that summarizes the last known global solution $\vect \alpha$. Also following~\cite{yang2013trading,ma2015adding} for faster convergence, each worker in our algorithm solves the following perturbed local dual problem, which we henceforth call the \emph{subproblem}:
\begin{align}
   \max_{\gvec{\delta}_{[k]} \in \mathbb{R}^n}  Q_{k}^{\sigma}(\gvec{\delta}_{[k]}; \vect v, \gvec{\alpha}_{[k]}) := & {-} \frac{1}{n_k}\sum_{i \in \mathcal{I}_k} \phi^*({-}\alpha_i {-} \delta_i) {-} \frac{\lambda}{S} g^*(\vect v) \nonumber \\
  & - \langle \frac{1}{n} X^{\top}_{[k]} \nabla g^*(\vect v), \gvec{\delta}_{[k]} \rangle - \frac{\lambda \sigma}{2} \norm{\frac{1}{\lambda n}\matrx X_{[k]}\vect \delta_{[k]}}^2  
\label{equ:local_opt}
\end{align}
where $\gvec{\delta}_{[k]}$ denotes the local (incremental) update to the dual variable $\vect \alpha_{[k]}$ and the \emph{scaling parameter} $\sigma$ measures the difficulty of solving the given data partition~(see~\cite{yang2013trading, ma2015adding}) and must be chosen such that
\begin{equation}
\label{equ:sigma_bound}
\sigma\ge \sigma_{min}:=\nu\max_{\vect\alpha\in \mathbb{R}^n}\frac{\norm{\matrx X\vect\alpha}^2}{\sum^K_{k=1}\norm{\matrx X\vect\alpha_{[k]}}^2}
\end{equation}
where the {\em aggregation parameter} $\nu\in [\frac{1}{S},1]$ is the weight given by the master to each of local updates from the contributing workers while computing the global update. The second term in the objective of our subproblem has denominator $S$ in stead of $K$.
Unlike the synchronous all reduce approach in~\cite{ma2015adding}, our asynchronous method merges the local updates from only $S$ out of $K$ nodes in each global update. By Lemma~3.2 in~\cite{ma2015adding}, $\sigma:=\nu S$ is a safe choice to hold condition~\eqref{equ:sigma_bound}.

\subsection{Asynchronous updates by cores in a worker node}

\begin{algorithm}[t]
   \caption{Hybrid-DCA: Worker node $k$}
   \label{alg:SharedMem}
  \KwIn{Initial $\gvec{\alpha_{[k]}} \in \mathbb{R}^n$, data partition $I_k$, \newline scaling parameter $\sigma$, aggregation parameter $\nu$}
$\bvec{v} \leftarrow \frac{1}{\lambda n} X \gvec{\alpha}_{[k]}$\;
\For{$t \leftarrow 0, 1, \ldots$}{
$\gvec{\delta}_{[k]} \leftarrow \bvec{0}, \quad \bvec{v}_{old} \leftarrow \bvec{v}$\;
  \For{cores $r \leftarrow 1, \ldots, R$ \textbf{in parallel}} {
     \For{$h \leftarrow 0, 1, \ldots, H-1$} {
       Randomly pick $i$ from $I_{k,r}$\;
       $\varepsilon \gets \argmax_{\varepsilon} Q^{\sigma}_{k}(\varepsilon \vect e_i; \bvec{v}, \gvec{\alpha}_{[k]}+\gvec{\delta}_{[k]})$\;
       $\vect \delta_{[k]} \leftarrow  \vect \delta_{[k]} + \varepsilon \vect e_i$\;
       $\bvec{v} \xleftarrow{\text{\em atomic}} \bvec{v} + \nabla g^*\left(\frac{1}{\lambda n} X \varepsilon \vect e_i\right)$\;
    }
  }
  send $\Delta \vect v \gets \vect v - \vect v_{old}$ to the master\;
  receive $\vect v$ from the master\;
  $\gvec{\alpha}_{[k]} \leftarrow \gvec{\alpha}_{[k]} + \nu \gvec{\delta}_{[k]}$\;
}
\end{algorithm}

In each communication round, each worker $k$ solves its subproblem using a parallel asynchronous DCA method~\cite{hsieh2015passcode} on the $R$ cores. Let the data partition $\mathcal{I}_k$ stored in the shared memory be logically divided into $R$ subparts where subpart $\mathcal{I}_{k,r} \subseteq \mathcal{I}_k,r=1,\ldots,R$, is exclusively used by core $r$. In each of the $H$ iterations, core $r$ chooses a random coordinate $i \in \mathcal{I}_{k,r}$ and updates $\vect \delta_{[k]}$ in the $i$th unit direction by a step size $\varepsilon$ computed using a single variable optimization problem:
\begin{align}
  \varepsilon = \argmax_{\varepsilon \in \mathbb{R}} Q_{k}^{\sigma}(\varepsilon \vect e_i; \bvec{v}, \gvec{\alpha}_{[k]} + \vect \delta_{[k]}) 
  \label{eq:dca_update}
\end{align}
which has a closed form solution for SVM problems~\cite{fan2008liblinear}, and a solution using an iterative solver for logistic regression problems~\cite{yu2011dual}. The local updates to $\vect v$ are also maintained appropriately. While the coordinates used by any two cores and hence the corresponding updates to $\vect \delta_{[k]}$ are independent of each other, there might be conflicts in the updates to $\vect v$ if the corresponding columns in $X$ have nonzero values in some common row. We use lock-free \emph{atomic} memory updates to handle such conflicts.
When all cores complete $H$ iterations, worker $k$ sends the accumulated update $\Delta \vect v$ from the current round to the master; waits until it receives the globally updated $\vect v$ from the master; and repeats for another round unless the master indicates termination.  

\subsection{Merging updates from workers by master}

If the master had to wait for the updates from all the workers, it could compute the global updates only after the slowest worker finished. To avoid this problem, we use \emph{bounded barrier}: in each round, the master waits for updates from only a subset $\mathcal{P}_S$ of $S \le K$ workers, and sends them back the global update $\vect v = \vect v + \nu \sum_{k \in \mathcal{P}_S} \Delta \vect v_k$. 
However, due to this relaxation, there might be some slow workers with out-of-date $\vect v$. When updates from such workers are merged by the master, it may degrade the quality of the global solution and hence may cause slow convergence or even divergence. We ensure sufficient freshness of the updates using \emph{bounded delay}: the master makes sure that no worker has a stale update older than $\Gamma$ rounds.
This asynchronous approach has two benefits: 1) the overall progress is no more bottlenecked by the slowest processor, and 2) the total number of communications is reduced. On the flip side, convergence may get slowed down for very small $S$ or very large $\Gamma$.

\begin{algorithm}[tb]
   \caption{Hybrid-DCA: Master node}
   \label{alg:hybrid}
   \SetKw{KwOr}{or}
   \KwIn{Initial $\gvec{\alpha} \in \mathbb{R}^n$, aggregation parameter $\nu$, \newline barrier bound $S$, delay bound $\Gamma$}
  $\bvec{v}^{(0)} \leftarrow \frac{1}{\lambda n}\matrx X \gvec{\alpha}$; \qquad $\mathcal{P} = \emptyset$\;
  \For{$t \leftarrow 0, 1, \ldots$}{
  \While{$|\mathcal{P}| < S$ \KwOr $\max_{k} \Gamma_k > \Gamma$}{
    receive update $\Delta \bvec{v}_k$ from some worker $k$\;
    $\mathcal{P} \gets \mathcal{P} \cup \{k\}$; \qquad $\Gamma_k \leftarrow 1$\;
  }
    $\mathcal{P}^{(t)}_S \gets S$ workers in $\mathcal{P}$ with oldest updates\;
    $\bvec{v}^{(t+1)} \leftarrow \bvec{v}^{(t)} + \nu \sum_{k \in \mathcal{P}^{(t)}_S} \Delta \bvec{v}_{k}$; \quad 
    $\mathcal{P} \gets \mathcal{P} \setminus \mathcal{P}^{(t)}_S$\;
    \lForEach{$k \notin \mathcal{P}^{(t)}_S$}{$\Gamma_k \leftarrow \Gamma_k + 1$}
    broadcast $\vect v^{(t+1)}$ to all workers in $\mathcal{P}^{(t)}_S$\;
  }
\end{algorithm}

\begin{figure*}[t]
  \centering
    \includegraphics[width=0.9\linewidth]{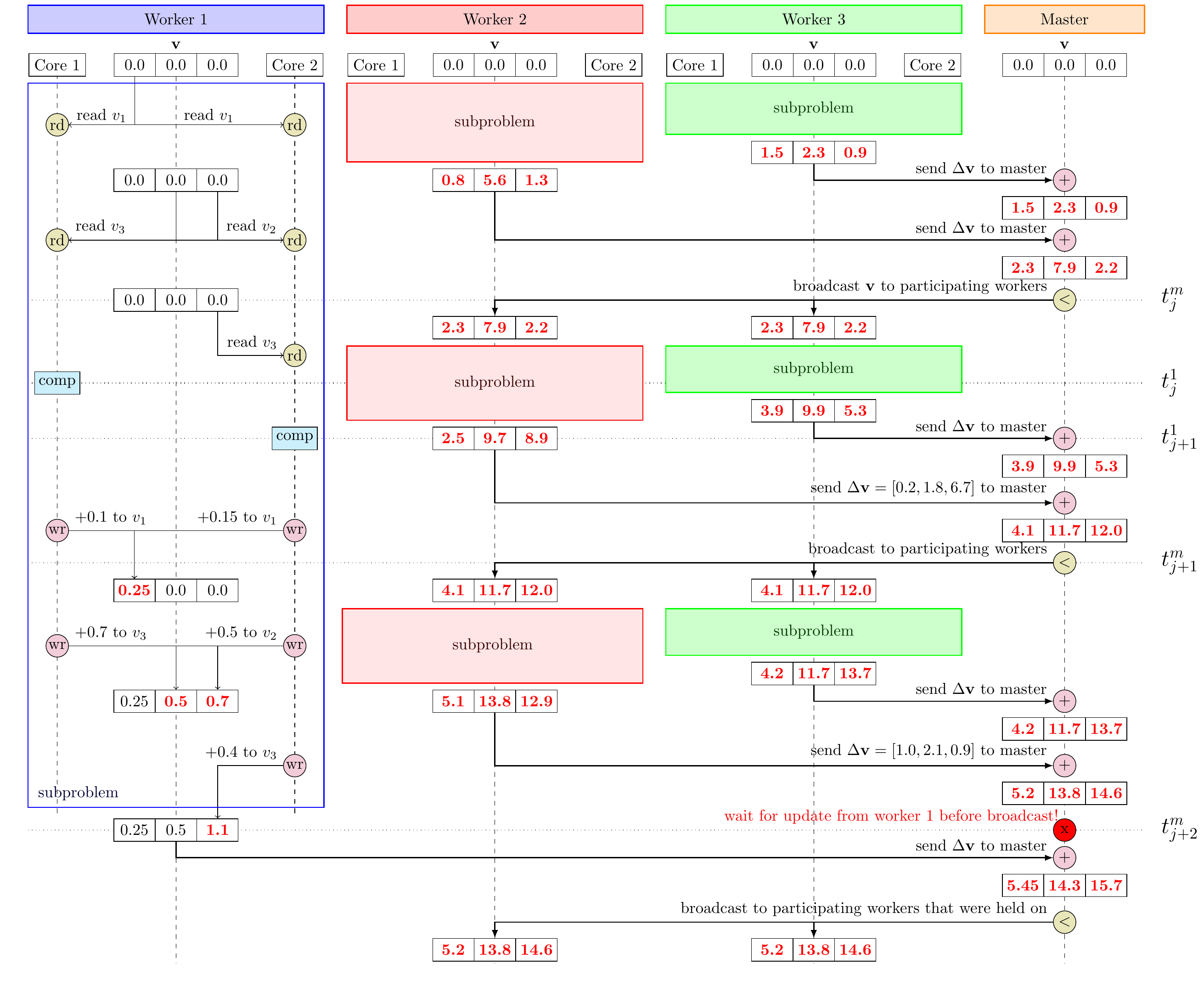}
  \caption{Sequence of important events in an example run of Hybrid-DCA where $n=12, d=3, K=3, R=2, S=2, \Gamma=2, \nu=1$.}
  \label{fig:seqdiag}
\end{figure*}

\vspace{5pt}\noindent{\bf Example:} Figure~\ref{fig:seqdiag} shows a possible sequence of important events in a run of our algorithm on a dataset having $n=12$ data points in $d=3$ dimensions using $K=3$ nodes each having $R=2$ cores such that each core works with only $|I_{k,r}|=2$ data points. The activities in solving the subproblem using $H=1$ local iterations in a round is shown in a rectangular box. For the first subproblem, core~1 and core~2 in worker~1 randomly selects dual coordinates such that the corresponding data points have nonzero entries in the dimensions $\{1,3\}, \{1,2,3\}$, respectively.  Each core first reads the entries of $\vect v$ corresponding to these nonzero data dimensions, and then computes the updates $[0.1, 0, 0.7], [0.15, 0.5, 0.4]$, respectively, and finally applies these updates to $\vect v$. The atomic memory updates ensure that all the conflicting writes to $\vect v$, such as $v_1$ in the first write-cycle, happen completely. At the end of $H$ local iterations by each core, worker~1 sends $\Delta \vect v = [0.25, 0.5, 1.1]$ to the master, the responsibility of which is shared by one of the $3$ nodes, but shown separately in the figure. By this time, the faster workers~2 and~3 already complete 3 rounds. As $S=2$, the master takes first 2 updates from $\mathcal{P}^{(1)}_S = \mathcal{P}^{(2)}_S = \{2, 3\}$ and computes the global updates using $\nu = 1$. However, as $\Gamma=2$, the master holds back the third updates from workers~2,3 until the first update from worker~1 reaches~master.  The subsequent events in the run are omitted in the figure.  

\section{Convergence Analysis}
\label{sec:analysis}

In this section we prove the convergence of the global solution computed by our hybrid algorithm. For ease we prove for $g(w) = \norm{w}^2$; the proof can be similarly extended for other regularizers $g(w)$. The analysis is divided into two parts. First we show that the solution of the subproblem computed by each node locally is indeed not far from the optimum. Using this result on the subproblem, we next show the convergence of the global solution. Though our proofs for the two parts are based on the works~\cite{hsieh2015passcode} and~\cite{ma2015distributed}, respectively, we need to make significant adjustments in the proofs due to our modified framework handling two cascaded levels of asynchronous updates. Here we outline the modifications, the complete proofs are given in the appendix.

\subsection{Near optimality of the solution to the local subproblem}

\begin{definition}
  For given $\vect v, \gvec{\alpha}_{[k]}$, a solution $\vect \delta_{[k]}$ to the subproblem~\eqref{equ:local_opt} is said to be $\Theta$-approximate, $\Theta \in [0,1)$, if 
  \begin{align}
     \mathbb{E}\left[ Q_{k}^{\sigma}(\gvec{\delta}^*_{[k]}; \vect v, \gvec{\alpha}_{[k]}) - Q_{k}^{\sigma}(\gvec{\delta}_{[k]}; \vect v, \gvec{\alpha}_{[k]}) \right] 
      \le \Theta \left[ Q_{k}^{\sigma}(\gvec{\delta}^*_{[k]}; \vect v, \gvec{\alpha}_{[k]}) - Q_{k}^{\sigma}(\bvec{0}; \vect v, \gvec{\alpha}_{[k]}) \right]
     \label{eq:subopt}
  \end{align}
  where $\vect \delta^*_{[k]}$ is the optimum solution to~\eqref{equ:local_opt}.
\end{definition}

The main challenge in using the results of~\cite{hsieh2015passcode} to prove~\eqref{eq:subopt} for the solution returned by the parallel asynchronous stochastic DCA solver used by each worker in Algorithm~\ref{alg:SharedMem} is to tackle the following changes: 1) the solver here solves only a part of the dual problem and 2) the subproblem is now perturbed (see Section~\ref{sec:algos}). While the first change is simply handled by considering the updates by the cores in worker~$k$ only, the second change needs modifications in each step of the proof in~\cite{hsieh2015passcode}, including the definition of the proximal operators $T_i$ and the assumptions. Here we rewrite the statements of the modified lemmas and assumptions.  

We consider the updates made in the current round by all the cores in the ascending order of the actual time point ($U_j^k$ in Figure~\ref{fig:seqdiag}) when the step size $\varepsilon$ of the update is computed (breaking ties arbitrarily) and prove~\eqref{eq:subopt} by showing sufficient progress in between two successive updates in this order, however, under some assumptions similar to those used in~\cite{hsieh2015passcode}.

Because of the atomic updates, the step size computation may not include all the latest updates, however, we assume all the updates before the ($j-\gamma$)-th update have already been written into $\vect v$. Let $\bar{\matrx X}_{[k]}$ denote the normalized data matrix where each row is $\bar{\vect x}_i^\top={\vect x_i^\top}/{\|\vect x_i\|}, i\in \mathcal{I}_k$. Define $M_{[k],i} = \max_{ \mathcal{D} \subseteq[d]} \| \sum_{t\in \mathcal{D}}\bar{\matrx X}_{[k](:,t)}X_{[k](i,t)}\|$, $M=\max_k \max_i M_{[k],i}$, where $[d]$ is the set of all the feature indices, and $\bar{\matrx X}_{[k](:,t)}$ is the $t$-th column of $\bar{\matrx X}_{[k]}$. Moreover, $R_{min}$ is defined as the minimum value of global data matrix, \ie $R_{min}=\min_{i=1,\dots,n}\|\vect x_i\|^2$.

\begin{assumption}[Bounded delay of updates $\gamma$]
	\begin{equation}
	\label{equ:gamma}
  (\gamma+1)^2 \le \frac{\sqrt{n_k}}{6eM}, \text{ where $e$ is the Euler's number.}
	\end{equation}
  \label{as:bounded_delay}
\end{assumption}
\vspace{-15pt}

\begin{definition}[Global error bound]
  For a convex function $f:\mathbb{R}^n \rightarrow \mathbb{R}$, the optimization problem: $\min_{\vect \beta} f(\vect \beta)$ admits a global error bound if there is a constant $\kappa$ such that 
\begin{equation}
\label{equ:global_error_bound}
\norm{\vect \beta-P_S(\vect \beta)}\le \kappa \norm{T(\vect \beta)-\vect \beta},
\end{equation}
where $P_S(\cdot)$ is the Euclidean projection to the set of optimal solutions, and $T:\mathbb{R}^{n}\rightarrow\mathbb{R}^{n}$ is the operator defined as
\begin{equation*}
T_i(\vect \beta)=\arg\min_u f\left(\vect \beta+(u-\beta_i)\vect e_t\right)\;\;\forall i\in [n].
\end{equation*}
The optimization problem admits a relaxed condition called global error bound from the beginning if \eqref{equ:global_error_bound} holds for any $\vect \beta$ satisfying $f(\vect \beta) \le F$ for some constant $F$.
\end{definition}

\begin{assumption}
  The local subproblem formulation~\eqref{equ:local_opt} admits global error bound from the beginning for $F=Q(\vect \delta^{(j)}_{[k]}; \vect v^{(j)}, \vect \alpha^{(j)}_{[k]})$ and any update $j$.
\end{assumption}
\vspace{-5pt}

The global error bound helps prove that our subproblem solver achieves significant improvement after each update. It has been shown that when the loss functions are hinge loss or squared hinge loss, the local subproblem formulation~\eqref{equ:local_opt} does indeed satisfy global error bound condition~\cite{hsieh2015passcode}.

\begin{assumption}
  The local subproblem objective~\eqref{equ:local_opt} is $L_{max}$-Lipschitz continuous. 
\end{assumption}
\vspace{-5pt}

\begin{assumption}[Bounded $M,L_{max}$]
\begin{equation*}
2L_{max}\left(1+\frac{e^2\gamma^2 M^2}{\sigma^2n_k}\right)\left(\frac{e^2\gamma^2M^2}{\sigma^2n_k}\right) \le 1
\end{equation*}
\label{as:bounded_m}
\end{assumption}
\vspace{-5pt}

\begin{lemma}
\label{thm:Theta}
When Assumptions~\ref{as:bounded_delay}-\ref{as:bounded_m} hold, the solutions computed in two successive updates by the {\em local subproblem} solver has a linear convergence rate in expectation, \ie
\begin{align*}
\label{equ:loacal_convergence_rate}
\mathbb{E}&\left[ Q_{k}^{\sigma}(\gvec{\delta}^*_{[k]}) - Q_{k}^{\sigma}(\gvec{\delta}^{(j)}_{[k]}) \right]  \nonumber \le \eta \left[ Q_{k}^{\sigma}(\gvec{\delta}^*_{[k]}) - Q_{k}^{\sigma}(\gvec{\delta}^{(j-1)}_{[k]}) \right]
\end{align*}
where $\gvec{\delta}^{(j)}_{[k]}$ is the $\vect \delta_{[k]}$ after the $j$th update, 
\begin{equation*}
\label{equ:eta}
\eta=1-\frac{\kappa R_{min}}{2nL_{max}}\left(1{-}\frac{2L_{max}}{R_{min}}\left(1{+}\frac{e^2\gamma^2 M^2}{\sigma^2\tilde{n}}\right)\left(\frac{e^2\gamma^2M^2}{\sigma^2\tilde{n}}\right)\right),
\end{equation*}
and $\tilde{n}=\max_k n_k$ is the size of the largest data part. Moreover, $\vect \delta^{(H)}_{[k]}$ is a $\Theta$-approximate solution for
\begin{equation}
\Theta = \eta^H.
\label{eq:theta}
\end{equation}
\end{lemma}
\vspace{-10pt}

\subsection{Convergence of global solution}

Although we showed that the local subproblem solver outputs a $\Theta$-approximate solution, we cannot directly apply the results of~\cite{ma2015distributed} for the global solution because our algorithm uses updates from only a subset of workers which is unlike the synchronous all-reduce of the updates from all workers used in~\cite{ma2015distributed}. We need to handle this asynchronous nature of the global updates, just like we handled asynchronous updates for the local subproblem.  Let us consider the global updates in the order the  master computed them (at time $U_t^m$ in Figure~\ref{fig:seqdiag}). Let $\vect \alpha^{(t)}$ denote the value of $\vect \alpha$ distributed across all the nodes at the time master computed $t$th global update $\vect v^{(t)}$. If $k \in \mathcal{P}^{(t)}_S$ then the update $\vect \delta^{(t)}_{[k]}$ has already been included in $\vect v^{(t)}$. However, if $k \notin \mathcal{P}^{(t)}_S$ then it may not be included. Let $\xi$ be such that for all $l \le \xi$ and for all $k$, $\vect \delta^{(l)}_{[k]}$ has been included in $\vect v^{(t)}$. By the design of our algorithm, $t-\Gamma\le\xi\le t-1$. Let $\hat{\vect \alpha}^{(t)}$ be defined as follows: $\hat{\vect \alpha}^{(t)}_{[k]}=\vect \alpha^{(t)}_{[k]}, \forall k\in \mathcal{P}^{(t)}_S$ and $=\vect \alpha^{(\xi)}_{[k]}$ for the latest $\xi$ for which the update is already included in global $\vect v$, $\forall k\notin \mathcal{P}^{(t)}_S$. Let $\vect w^{(t)},\hat{\vect w}^{(t)}$ be $\vect w(\vect \alpha^{(t)})$ and $\vect w(\hat{\vect \alpha}^{(t)})$ respectively. Note that $\vect w^{(t)}=\hat{\vect w}^{(t)}+\frac{1}{\lambda n}\sum_{l=\xi}^{t-1}\matrx X\vect \delta^{(l)}$.

\begin{lemma}
	\label{thm:D_alpha+1}
  For any dual $\vect \alpha^{(t)},\vect \delta^{(t)}\in \mathbb{R}^n$, primal $\hat{\vect w}^{(t)}=\vect w(\hat{\vect \alpha}^{(t)})$ and real values $\nu,\sigma$ satisfying (\ref{equ:sigma_bound}), it holds that
	\begin{align}
	\label{equ:D_j+1}
	D & \left( \vect \alpha^{(t)}+\nu\sum_{k\in\mathcal{P}^{(t)}_S}\vect \delta^{(t)}_{[k]}\right)\ge (1-\nu)D(\hat{\vect \alpha}^{(t)})
 -\frac{\lambda}{2}(\norm{\vect w^{(t)}}^2-\norm{\hat{\vect w}^{(t)}}^2) \nonumber\\
 +&\nu\sum_{k\in \mathcal{P}^{(t)}_S}Q^\sigma_k\left(\vect \delta^{(t)}_{[k]};\hat{\vect w}^{(t)}, \vect \alpha^{(t)}_{[k]}\right)
 -\frac{\nu}{n}\sum_{k\in \mathcal{P}^{(t)}_S}(\vect w^{(t)}-\hat{\vect w}^{(t)})^\top\matrx X\vect \delta^{(t)}_{[k]}. 
	\end{align}
\end{lemma}


\begin{assumption}
  There exists a $\varrho < e^{\frac{2}{\Gamma+1}}$ such that
	\begin{align}
	\label{equ:varrho_bound}
  \norm{\vect \delta^{(t-1)}}^2\le  \varrho\norm{\vect \delta^{(t)}}^2.
	\end{align}
\label{as:bounded_rho}
\end{assumption}
\vspace{-5pt}

\begin{lemma}
	\label{lem:global_convergence}
	If $\phi^*_i$ are all $(1/\mu)$-strongly convex and Assumptions~\ref{as:bounded_delay}-\ref{as:bounded_rho} are satisfied then for any $s\in [0,1]$, any round $t$ of Algorithm~\ref{alg:hybrid} satisfies
	\begin{equation}
    \mathbb{E}[D(\vect \alpha^{(t+1)})-D(\vect \alpha^{(t)})]\ge \Psi(1-\Theta)\left(sG(\hat{\vect \alpha})-\frac{\sigma}{2\lambda}\left(\frac{s}{n}\right)^2\hat{R}\right)
	\end{equation}
	where 
	\begin{align}
	\label{equ:Psi}
  \Psi:=&\nu\left(1-\frac{\Gamma^2 eML_{max}}{4\lambda n^2}-\frac{SML_{max}}{4n}\right)\le 1, \; \text{and}\\
	\label{equ:R}
	\hat{R}:=& -\frac{\lambda\mu n(1-s)}{\sigma s}\norm{\hat{\vect u}-\hat{\vect \alpha}}^2+\sum_{k\in\mathcal{P}_S}\norm{\matrx X(\hat{\vect u}-\hat{\vect \alpha})_{[k]}}^2,
	\end{align}
	for $\hat{\vect u}\in \mathbb{R}^n$ with $-\hat{u}_i\in\partial\phi_i(\vect w(\hat{\vect \alpha})^\top\vect x_i)$.
\end{lemma} 

Using the main results in~\cite{ma2015adding} and combining Lemma~\ref{thm:Theta} with Lemma~\ref{lem:global_convergence} gives us the following two convergence results, one for smooth loss functions and the other for the Lipschitz continuous loss functions. The theorems use the quantities $\sigma_{max} = \max_{k} \sigma_k, \; \sigma_{sum} = \sum_{k} \sigma_k n_k$ where $\forall k, \sigma_k = \max_{\alpha_{[k]} \in \mathbb{R}^n} \norm{\matrx X\vect \alpha_{[k]}}^2/\norm{\vect \alpha_{[k]}}^2$.  

\begin{theorem}
	If the loss functions $\phi_i$ are all $(1/\mu)$-smooth, then in $T_1$ iterations Algorithm~\ref{alg:hybrid} finds a solution with objective atmost $\epsilon_D$ from the optimal, \ie $\mathbb{E}[D(\alpha^*) - D(\alpha^{(T_1)})] \le \epsilon_D$ whenever $T_1 \ge C_1 \log{\frac{1}{\epsilon_D}}$ where $C_1 = \frac{1}{\Psi(1-\Theta)} ( 1 + \frac{\sigma_{max} \sigma}{\nu \lambda n})$ and $\Theta$ is given by~\eqref{eq:theta}. Furthermore, in $T_2$ iterations, it finds a solution with duality gap atmost $\epsilon_{gap}$, \ie  $\mathbb{E}[P(\vect{w}(\alpha^{(T_2)})) - D(\alpha^{(T_2)})] \le \epsilon_{gap}$ whenever $T_2 \ge C_1 \log{\frac{C_1}{\epsilon_D}}$.
\label{thm:convex}
\end{theorem}

\begin{theorem}
	If the loss functions $\phi_i$ are all $L$-Lipschitz, then in $T_1$ iterations Algorithm~\ref{alg:hybrid} finds a solution with duality gap atmost $\epsilon_{gap}$, \ie  $\mathbb{E}[P(\vect{w}(\bar{\alpha})) - D(\bar{\alpha})] \le \epsilon_{gap}$ for the average iterate $\bar{\alpha} = \frac{1}{T_1-T_0} \sum_{t=T_0+1}^{T_1-1} \alpha^{(t)}$ whenever $T_1 \ge T_0 + \max\{\ceil{\frac{1}{\Psi(1-\Theta)}}, \frac{4L^2\sigma_{sum}\sigma}{\lambda n^2 \epsilon_{gap} \Psi(1-\Theta)}\}$, and $T_0 \ge \max\{0, \ceil{\frac{1}{\Psi(1-\Theta)}\log\frac{2\lambda n^2(D(\alpha^{*} - D(\alpha^{(0)})}{4L^2\sigma_{sum}\sigma}}\} + \max\{0, \frac{2}{\Psi(1-\Theta)}(\frac{8L^2\sigma_{sum}\sigma}{\lambda n^2 \epsilon_{gap}} -1)\}$ and $\Theta$ is given by~\eqref{eq:theta}.
\label{thm:lipschitz}
\end{theorem} 

Theorem~\ref{thm:lipschitz} establishes the convergence for $L$-Lipschitz continuous loss functions, and Theorem~\ref{thm:convex} proves a linear convergence rate for smooth convex loss functions.

\section{Communication Cost Analysis}

In each communication round, the algorithms based on synchronous updates on all $K$ nodes require $2K$ transmissions, each consisting of all values of $\vect v$ or $\Delta \vect v$. Half of these transmissions are from the workers to the master and the rest are from the master to the workers. Whereas, our asynchronous update scheme requires $2S$ transmissions in each round.

\section{Experimental Results}
\label{sec:results}

We implemented our algorithm in C++ where each node runs exactly one MPI process which in turn runs one OpenMP thread on each core available within the node and the main thread handles the inter-node communication. We evaluated for hinge loss, though it applies to other loss functions too, on four datasets from LIBSVM website as shown in Table~\ref{tab:datasets}, using upto 16 nodes available with the Hornet cluster at University of Connecticut where each node has 24 Xeon E5-2690 cores and 128 GB main memory. The last column in Table~\ref{tab:datasets} gives the total number of non-zero entries in the matrix $X$ for each of the four dataset we used.  

We experimented with the following algorithms: 1) \baseline: an implementation of DCA~\cite{hsieh2008dual}, 2) \cocoap~\cite{ma2015distributed}, 3) \passcode~\cite{hsieh2015passcode} and 4) our \hdca. The algorithm parameters were varied as follows: 1) regularization parameter $\lambda \in \{10^{-3}, 10^{-4}, 10^{-5}\}$, 2) local iterations $H=\{10000, 20000, 40000\}$, 3) aggregation parameters $\nu=1$, and 4) scaling parameter $\sigma = K,S$ for \cocoap, \hdca, respective, as recommended in~\cite{ma2015distributed}. For different combinations of $\lambda, H$, we observed similar patterns of results and report for $\lambda=10^{-4}, H=40000$ only. The details of other parameter values are given later. 

\begin{table}[t]
\setlength{\tabcolsep}{6pt}
\centering
\caption{Datasets}
\label{tab:datasets}
\begin{tabular}{lcrrrr}
\toprule
\textbf{Dataset} & \textbf{LIBSVM name} & $n$ & $d$ & $nnz$ & \textbf{File size}\\
\midrule
{\tt rcv1}       &  rcv1\_test & 677,399 &      47,236 & 49,556,258 & 1.2 GB\\
{\tt webspam}    &  webspam & 280,000 &  16,609,143 & 1,045,051,224 & 20 GB\\
{\tt kddb}       & kddb train     & 19,264,097 &  29,890,095 & 566,345,888 & 5.1 GB\\
{\tt splicesite} & splice\_site.t & 4,627,840 &  11,725,480 & 15,383,587,858 & 280 GB\\
\bottomrule
\end{tabular}
\end{table}

\subsection{Comparison with existing algorithms}
\begin{figure*}[tb]
\centering
\includegraphics[width=1.01\textwidth]{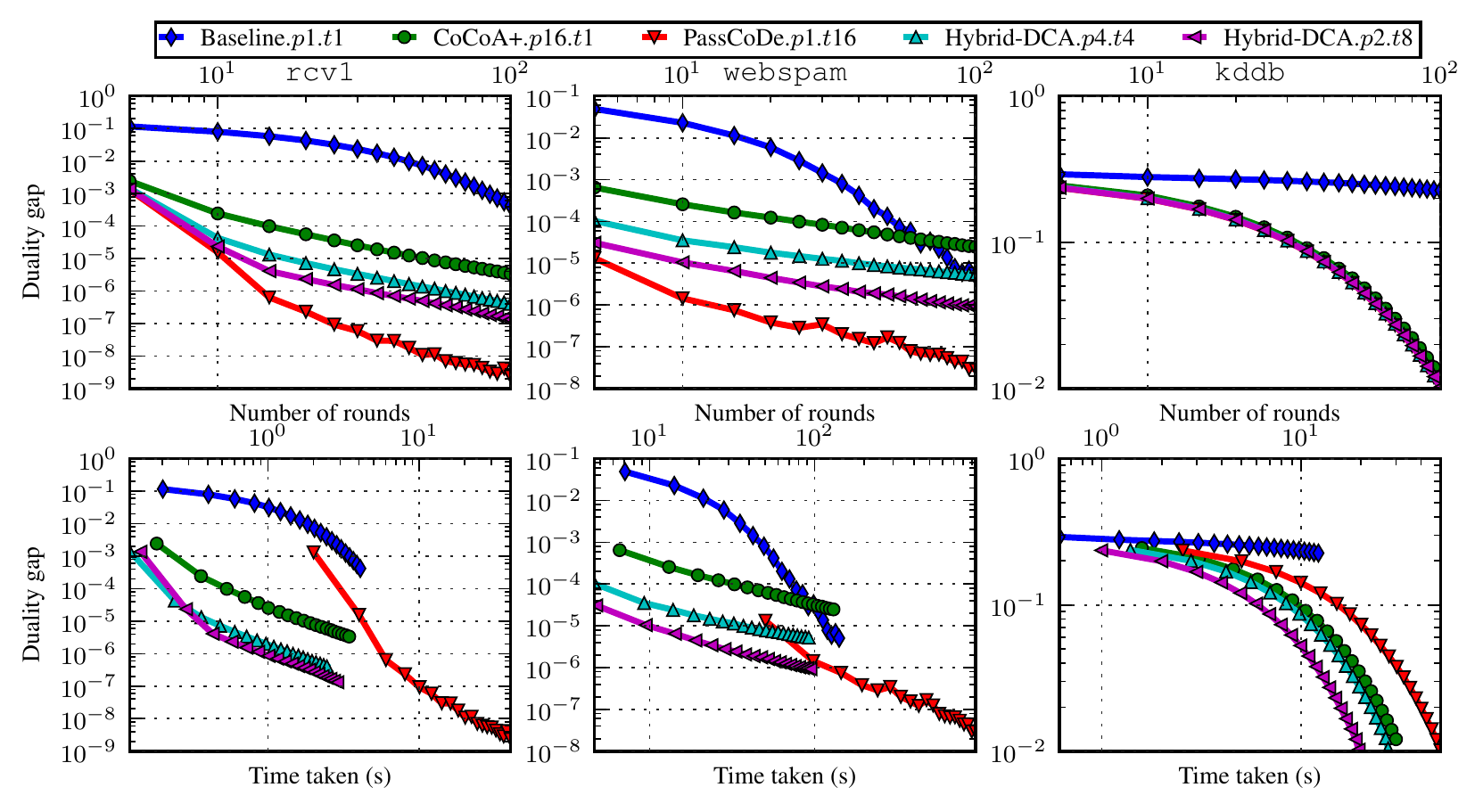}
\caption{Performance of different solvers on three datasets, {\tt rcv1} (left column), {\tt webspam} (middle column), and {\tt kddb} (right column), in terms of the progress of the duality gap across the number of rounds (top row) and across the wall time taken (bottom row).}
\label{fig:results}
\end{figure*}

Figure~\ref{fig:results} shows the progress of duality gap achieved by the four algorithms on three smaller datasets. We chose the number of nodes ($p \le K$) and the number of cores ($t \le R$) per node such that the total number of worker cores ($p \times t$) is the same (16) for all algorithms except \baseline. The duality gap is measured as $P(\vect v) - D(\vect \alpha)$ where $\vect v$ is the estimate of $\vect w(\vect \alpha)$ shared across the nodes. When $S<K$, it is not possible for the master in \hdca to gather the parts of $P(\vect v)$ from all workers at the end of each round. We let the master temporarily store $\vect v$ in disk after each round and at the end of all stipulated rounds, the workers compute the respective parts of $P(\vect v)$ from the stored $\vect v$ and the master computes the duality gap using a series of synchronous all-reduce computations from all the workers.  The bottom row shows the progress of the duality gap across time, while the top row shows progress across each round that  consists of a communication round in \cocoap and \hdca whereas consists of $H$ local updates in \baseline and \passcode. In this  experiment, \hdca uses $S=p$ and $\Gamma=1$ making the global updates synchronous. The progress of baseline is slow as it applies only $H$ updates compared to $H \times p \times t$ updates by the other algorithms. In terms of time, \hdca clearly outperforms both $\cocoap$, as expected, and $\passcode$ which incurs a larger number of cache-misses when many cores are used. In terms of the number of rounds, \passcode outperforms both \cocoap and \hdca, as expected. However, \passcode is not scalable beyond the number of cores in a single node. As the number of nodes increases, the convergence of \hdca becomes slower due to the costly merging process of many distributed updates.

\subsection{Speedup}

\begin{figure*}[tb]
\centering
\includegraphics[width=0.99\textwidth]{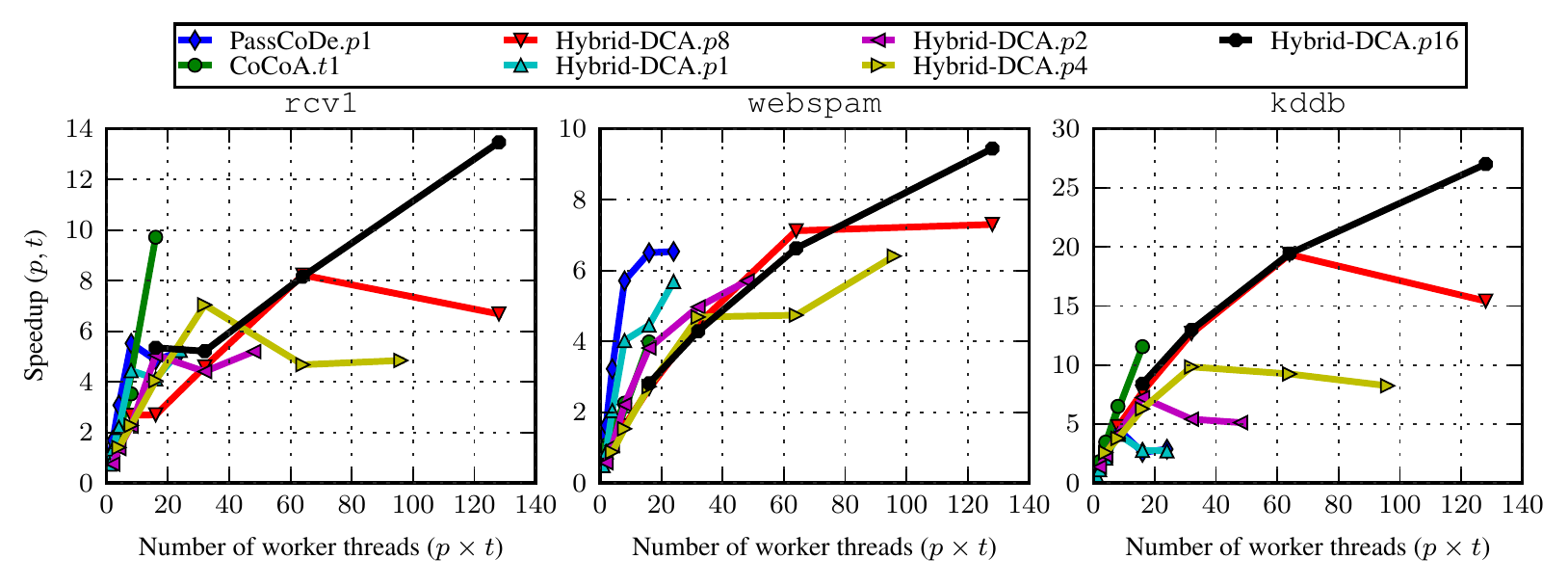}
\caption{Speedup of different parallel or distributed solvers with respect to the sequential implementation \baseline.}
\label{fig:speedup}
\end{figure*}

We ran sufficient rounds of each of the four algorithms such that the duality gap falls below a threshold and noted the time taken by the algorithms to achieve this threshold. Figure~\ref{fig:speedup} shows the speedup$(p,t)$ of all the algorithms except, \baseline, computed as the ratio of the time taken by \baseline to the time taken by the target algorithm on $p$ nodes each with $t$ cores. The thresholds used were $10^{-4}, 10^{-5}, 10^{-1}$ for {\tt rcv1, webspam, kddb}, respectively. \passcode can be run only on a single node; so we vary only the number of cores. \cocoap uses only 1 core per node. We ran \cocoap and \hdca on $p \in \{2, 4, 8, 16\}$ nodes and plotted them separately. For each $p$, \hdca uses $t\in\{2, 4, 8, 16, 24\}$ cores per node, however, under the restriction that the total number of worker cores ($p \times t$) does not exceed 144, a limit set by our HPC usage policy. When $t>8$, the number of cache-misses increases due to thread switching on the physical cores and reduces speedup for both \passcode and \hdca. This could be improved by carefully scheduling the OpenMP threads to the same physical cores. Nevertheless, \hdca has good speedup for $t \le 8$, as evident for $p=16$. 

\subsection{Effects of the parameter $S$}

\begin{figure*}[tb]
\centering
\includegraphics[width=0.99\textwidth]{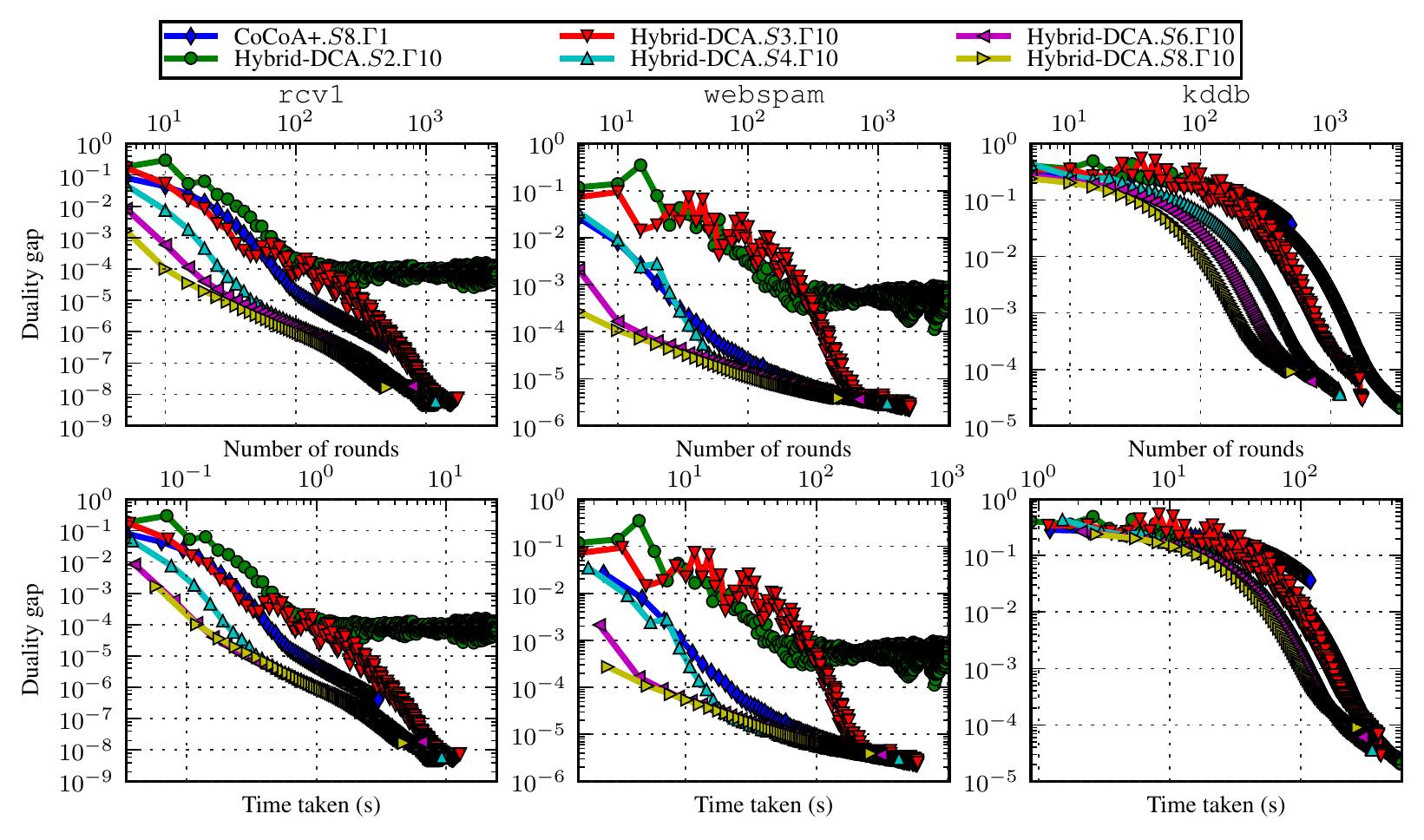}
\caption{Effect of varying $S$ on $p=8$ worker nodes, with $\Gamma$ fixed at 10.}
\label{fig:vary_s}
\end{figure*}

Figure~\ref{fig:vary_s} shows the results of varying $S \in \{2,3,4,6,8\}$ with fixed $\Gamma=10$ on $p=8$ nodes each with $t=8$ cores. When $S < p/2$, only a minority of the workers contribute in a round and the duality gap does not progress below some certain level. On the other hand, when at least half of the workers contribute in each round, it is possible to achieve the same duality level obtained using all the workers. However, the reduction in time per round is eventually eaten by the larger number of rounds required to achieve the same duality gap. Nevertheless, the approach is useful for HPC platforms with heterogeneous nodes, unlike ours, where the waiting for updates from all workers has larger penalty per round, or for the case, where the need is to run for a specified number of rounds and quickly achieve a reasonably good duality gap.

\subsection{Effects of the parameter $\Gamma$}

\begin{figure*}[tb]
\centering
\includegraphics[width=0.99\textwidth]{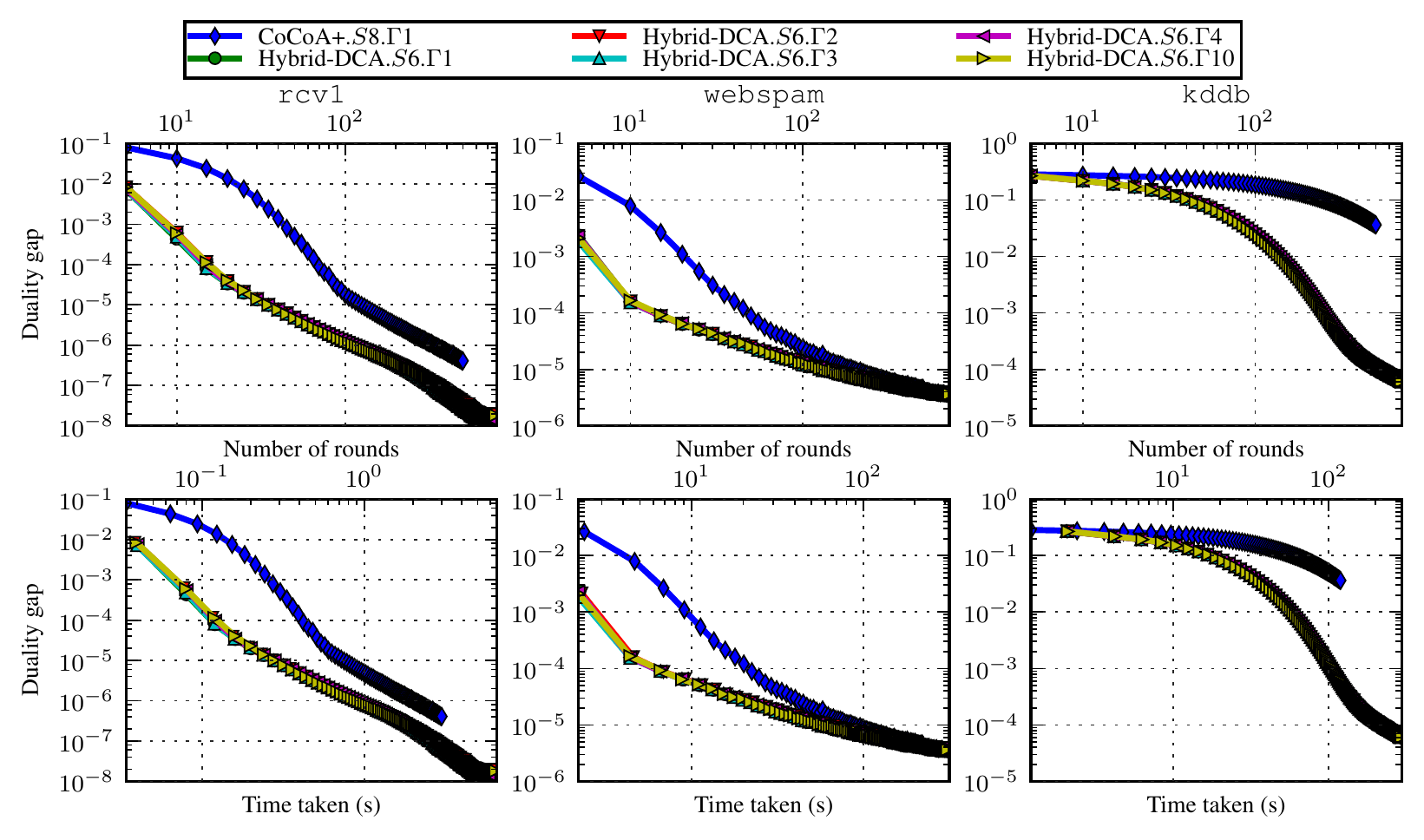}
\caption{Effect of varying $\Gamma$ on $p=8$ worker nodes, with $S$ fixed at 8.}
\label{fig:vary_tau}
\end{figure*}

Figure~\ref{fig:vary_tau} shows the results of varying $\Gamma \in \{1,2,3,4,10\}$ with fixed $S=6$ on $p=8$ nodes each with $t=8$ cores. We do not see much effect of $\Gamma$ as the HPC platform used for our experiments has homogeneous nodes. Our experimentation showed that even if we use $\Gamma=10$, the stale value at any worker was for at most $4$ rounds. We expect to see a larger variance of staleness in case of heterogeneous nodes.

\subsection{Performance on a big dataset}

\begin{figure*}[tb]
\centering
\includegraphics[width=0.99\textwidth]{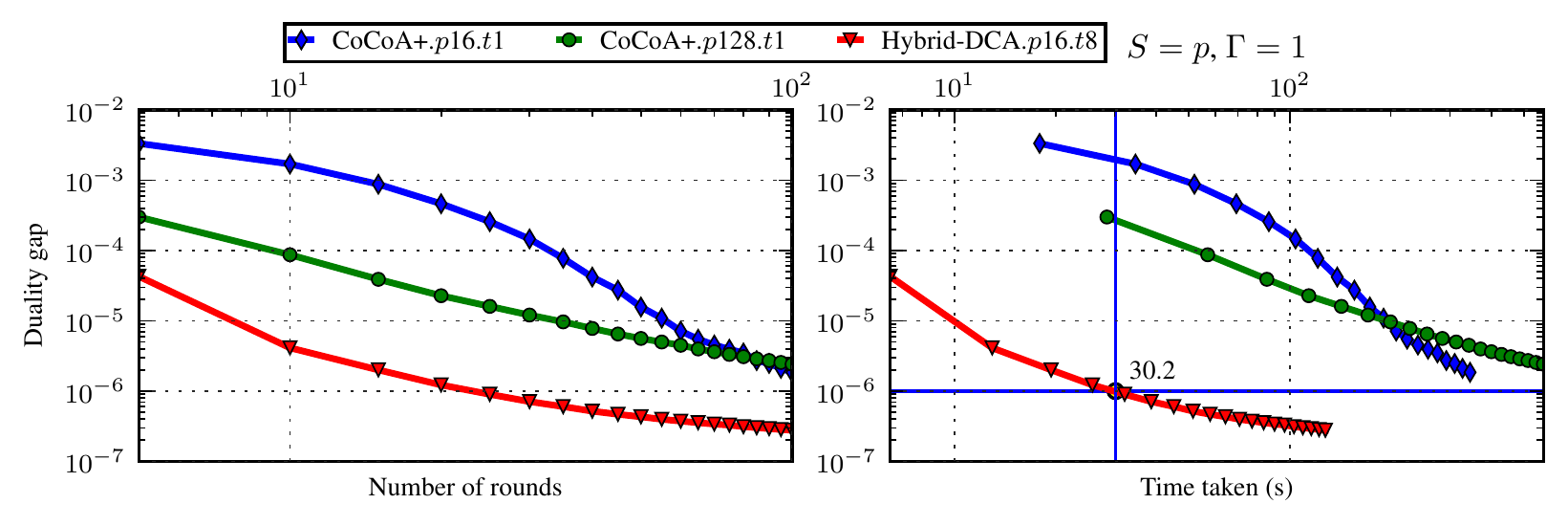}
\caption{Performance of \hdca on big dataset {\tt splicesite}.}
\label{fig:bigdataset}
\end{figure*}

We experimented our hybrid algorithm on the big dataset {\tt splicesite} of size about 280 GB and compared with the previous best algorithm \cocoap. Because of the enormous size, the dataset cannot be accommodated on a single node and hence \passcode cannot be run on this dataset. In this experiment, we used the number of local iterations $H=10,000$. The results are shown in Figure~\ref{fig:bigdataset} where the progress of duality gap across the rounds of communication is shown on the left and across the wall time on the right. To achieve a duality gap of at least $10^{-6}$ on 16 nodes, \cocoap took more than 300 seconds which somewhat matches the 1200 seconds (20 minutes) time on 4 nodes reported in~\cite{ma2015distributed}. \hdca on 16 nodes each using 8 cores took about 30 seconds to achieve the same duality gap, showing enough evidence about the scalability of our algorithm. One could also argue that \cocoap can be run on all these 16x8=128 cores, treating each core as a distributed node. We also experimented with this configuration which achieved better progress on duality gap than \cocoap on 16 nodes, however, still performed far worse than \hdca in terms of both the number of rounds and the time taken.


\section{Conclusions}
\label{sec:conclusions}
In this paper, we present a hybrid parallel and distributed asynchronous stochastic dual coordinate ascent algorithm utilizing modern HPC platforms with many nodes of multi-core shared-memory systems. We analyze the convergence properties of this novel algorithm which uses asynchronous updates at two cascading levels: inter-cores and inter-nodes. Experimental results show that our algorithm is faster than the state-of-the-art distributed algorithms and scales better than the state-of-the-art parallel algorithms.

{\small
\bibliographystyle{plainnat}
\bibliography{HybridDCA}
}

\newpage

\appendix
\noindent{\LARGE \bfseries \mytitle}
\section{Proof of near optimality of local subproblem solution}

\begin{definition}
First, we define that:
\begin{align*}
\beta^{l+1}_t=&\begin{cases}
T_t\left(\hat{\vect w}^l,\beta^l_t\right) & \text{if}~t=i(l),  \\
\beta^l_t & \text{if}~t\neq i(l)
\end{cases},& \varepsilon^l=&\beta^{l+1}_{i(l)}-\beta^{l}_{i(l)}, \\
\tilde{\beta}^{l+1}=& T\left(\hat{\vect w}^l,\vect \beta^l\right), &\bar{\beta}^{l+1}=& T\left(\bar{\vect w}^l,\vect \beta^l\right),
\end{align*}
where $\vect \beta^l$ denotes the sequence generated by the {\em local atomic solver} and $\hat{\vect w}^l$ denotes the actual values of $w$ maintained at update $l$ in the {\em local atomic solver}. Note that, $\tilde{\beta}^{l+1}_{i(l)}=\beta^{l+1}_{i(l)}$ and $\tilde{\beta}^{l+1}=\mathrm{prox}\left(\vect \beta^{l}-\frac{\lambda n}{\sigma}\bar{\matrx X}\hat{\vect w}^l\right)$.
\end{definition}

\begin{lemma}
\label{thm:lemma_alpha}
Under Assumption~\ref{as:bounded_delay}, and let $\vect \beta^l_{[k]}=\vect\alpha_{[k]}+\nu\vect\delta^l_{[k]}$, $\rho=\left(1+\frac{6(\gamma+1)eM}{\sqrt{n_k}}\right)^2$. Then, the local subproblem satisfy:
	\begin{equation}
	\label{equ:lemma_alpha}
	\mathbb{E}\left[\norm{\vect \beta_{[k]}^{l-1}-\tilde{\vect \beta}_{[k]}^{l}}^2\right] \le \rho \mathbb{E}\left[\norm{\vect \beta_{[k]}^{l}-\tilde{\vect \beta}_{[k]}^{l+1}}^2\right],
	\end{equation}
where $l\neq h$, represents the $l$-th update to $\omega$ in a {\em local solver}.
\end{lemma}

\subsection{Proof of Lemma \ref{thm:lemma_alpha}}

\begin{proof}
We omit the subscript $_{[k]}$ of the notations, which specifies the $k$-th data partition, in the proof. For all $i\in \mathcal{I}_k$, we have following definitions:
\begin{align*}
h_i(u):=&\frac{\phi^*_i(-u)}{n\norm{\vect x_i}^2}+\frac{\lambda}{2}\left(\frac{1}{S}-\frac{1}{\sigma}\right)\frac{\norm{\vect w}^2}{\norm{\vect x_i}^2} \\
\mathrm{prox}_i(s):=&\argmin_u \frac{1}{2}(u-s)^2+h_i(u)\\
T_i(\vect w,s):=&\argmax_u -\frac{1}{\sigma}\frac{\lambda}{2}\norm{\vect w}^2-\frac{1}{n}\vect w^\top\vect x_i(u-s)-\frac{\lambda}{2}\sigma(\frac{1}{\lambda n}\vect x_i(u-s))^2\\
&-\frac{1}{n}\phi^*_i(-u)-\frac{\lambda}{2}\left(\frac{1}{K}\norm{\vect w}^2-\frac{1}{\sigma}\norm{\vect w}^2\right) \\
=&\argmax_u -\frac{\lambda}{2}\norm{\frac{\vect w}{\sqrt{\sigma}}+\frac{\sqrt{\sigma}}{\lambda n}(u-s)\vect x_i}^2-\norm{\vect x_i}^2h_i(u) \\
=&\argmin_u \frac{1}{2}\left(u-\left(s-\frac{\lambda n\vect w^\top\vect x_i}{\sigma\norm{\vect x_i}^2}\right)\right)^2+h_i(u),
\end{align*} 
where $\vect w\in \mathbb{R}^d$ denotes any fixed vector and $s\in \mathbb{R}$. $\mathrm{prox}(s)$ denotes the proximal operator. We can see the connection of above operator and proximal operator: $T_i(\vect w,s)=\mathrm{prox}_i\left(s-\frac{\vect w^\top\vect x_i}{\sigma\norm{\vect x_i}^2}\right)$. Here both $h_i(u)$ and $T_i(\vect \wedge,s)$ were revised from \cite{hsieh2015passcode} to satisfy the subproblem \ref{equ:local_opt}.

\begin{proposition}
\label{pro:expectation}
\begin{equation}
\label{equ:alpha_expectation}
\mathbb{E}_{i(l)}\left(\norm{\vect \beta^{l+1}-\vect \beta^l}^2\right)=\frac{1}{n}\norm{\tilde{\vect \beta}^{l+1}-\vect\beta^l}^2,
\end{equation}
\end{proposition}
\begin{proposition}
\label{pro:M}
\begin{equation}
\label{equ:delta_alpha_bound}
\norm{\bar{\matrx X}\bar{\vect w}^j-\bar{\matrx X}\hat{\vect w}^j}\le \frac{1}{\lambda n}M \sum_{t=j-\gamma}^{j-1}|\varepsilon^t|,
\end{equation}
\end{proposition}
\begin{proposition}
\label{pro:T_w_s}
\begin{equation}
\label{equ:T_w_s}
|T_i\left(\vect w_1,s_1\right)-T_i\left(\vect w_2,s_2\right)|\le \left|s_1-s2+\frac{\left(\vect w_1-\vect w_2\right)^\top\vect x_i}{\norm{\vect x_i}^2}\right|,
\end{equation}
\end{proposition}
\begin{proposition}
\label{pro:rho}
Let $M\ge 1$, $q=\frac{6(\gamma+1)eM}{\sqrt{n}}$, $\rho=(1+q)^2$, and $\theta=\sum_{t=1}^\gamma \rho^{t/2}$. If $q(\gamma+1)\le 1$ and $\sigma\ge 1$, then $\rho^{(\gamma+1)/2}\le e$, and
\begin{equation}
\label{equ:rho}
\rho^{-1}\le 1-\frac{4}{\sqrt{n}}-\frac{4M+4M\theta}{\sqrt{n}}\le 1-\frac{4}{\sqrt{n}}-\frac{4M+4M\theta}{\sigma\sqrt{n}},
\end{equation}
\end{proposition}
\begin{proposition}
\label{pro:convex}
For all $j>0$, we have
\begin{align}
\label{equ:strong_convexity}
D\left(\vect \alpha^j\right)\le& D\left(\bar{\vect \alpha}^{j+1}\right)-\frac{\sigma\norm{\vect x_{i(j)}}^2}{2}\norm{\vect \alpha^j-\bar{\vect \alpha}^{j+1}}^2,\\
\label{equ:L_L_continuity}
D\left(\vect \alpha^{j}\right)\ge& D\left(\bar{\vect \alpha}^{j+1}\right)-\frac{L_{max}}{2}\norm{\vect \alpha^{j}-\bar{\vect \alpha}^{j+1}}^2
\end{align}
\end{proposition}
Above propositions are similar to \cite{hsieh2015passcode}. We keep the conclusions of those propositions for the future use in our proof. We prove Eq (\ref{equ:lemma_alpha}) by induction. As shown in \cite{hsieh2015passcode}, we have
\begin{equation}
\label{equ:alpha_inequality}
\norm{\vect \beta^{l-1}-\tilde{\vect \beta}^l}^2-\norm{\vect \beta^{l}-\tilde{\vect \beta}^{l+1}}^2\le 2\norm{\vect \beta^{l-1}-\tilde{\vect \beta}^l}\norm{\vect \beta^l-\tilde{\vect \beta}^{l+1}-\vect \beta^{l-1}+\tilde{\vect \beta}^l}.
\end{equation}
The second of factor in the r.h.s of Eq \ref{equ:alpha_inequality} is bounded as follows with the revisions:
\begin{align}
\label{equ:lemma_alpha_1}
&\norm{\vect \beta^l-\tilde{\vect \beta}^{l+1}-\vect \beta^{l-1}+\tilde{\vect \beta}^l}\nonumber\\
\le&\norm{\vect \beta^l-\vect \beta^{l-1}}+\norm{\mathrm{prox}\left(\vect \beta^l-\frac{\lambda n}{\sigma}\bar{\matrx X}\hat{\vect w}^{l}\right)-\mathrm{prox}\left(\vect \beta^{l-1}-\frac{\lambda n}{\sigma}\bar{\matrx X}\hat{\vect w}^{l-1}\right)}\nonumber\\
\le &\norm{\vect \beta^l-\vect \beta^{l-1}}+\norm{\left(\vect \beta^l-\frac{\lambda n}{\sigma}\bar{\matrx X}\hat{\vect w}^{l}\right)-\left(\vect \beta^{l-1}-\frac{\lambda n}{\sigma}\bar{\matrx X}\hat{\vect w}^{l-1}\right)}\nonumber\\
\le & 2\norm{\vect \beta^l-\vect \beta^{l-1}}+\frac{\lambda n}{\sigma}\norm{\bar{\matrx X}\hat{\vect w}^l-\bar{\matrx X}\hat{\vect w}^{l-1}\nonumber}\\
= & 2\norm{\vect \beta^l-\vect \beta^{l-1}}+\frac{\lambda n}{\sigma}\norm{\bar{\matrx X}\hat{\vect w}^l-\bar{\matrx X}\bar{\vect w}^l+\bar{\matrx X}\bar{\vect w}^l-\bar{\matrx X}\bar{\vect w}^{l-1}+\bar{\matrx X}\bar{\vect w}^{l-1}-\bar{\matrx X}\hat{\vect w}^{l-1}}\nonumber\\
\le & 2\norm{\vect \beta^l-\vect \beta^{l-1}}+\frac{\lambda n}{\sigma}\left(\norm{\bar{\matrx X}\bar{\vect w}^l-\bar{\matrx X}\bar{\vect w}^{l-1}}+\norm{\bar{\matrx X}\hat{\vect w}^l-\bar{\matrx X}\bar{\vect w}^{l}}+\norm{\bar{\matrx X}\bar{\vect w}^{l-1}-\bar{\matrx X}\hat{\vect w}^{l-1}}\right)\nonumber\\
\le &\left(2+2\frac{\lambda n}{\sigma}\frac{M}{\lambda n}\right)\norm{\vect \beta^l-\vect \beta^{l-1}}+2\frac{\lambda n}{\sigma}\frac{M}{\lambda n}\sum_{t=l-\gamma-1}^{l-2}|\varepsilon^t| \;\;\mathrm{(Proposition~\ref{pro:M})} \\
\le &\left(2+2\frac{M}{\sigma}\right)\norm{\vect \beta^l-\vect \beta^{l-1}}+2\frac{M}{\sigma}\sum_{t=l-\gamma-1}^{l-2}|\varepsilon^t|
\end{align}
No we start the induction. Although some steps may be the same as the steps in \cite{hsieh2015passcode}, we still keep them here to make the proof self-contained.

\textbf{Induction Hypothesis.} We prove the following equivalent statement. For all $j$,
\begin{equation*}
\mathbb{E}\left(\norm{\vect \beta^{l-1}-\tilde{\vect \beta}^l}^2\right)\le \rho \mathbb{E}\left(\norm{\vect \beta^{l}-\tilde{\vect \beta}^{l+1}}^2\right),
\end{equation*}
\textbf{Induction Basis.} When $l=1$,
\begin{align*}
\mathbb{E}\left(\norm{\vect \beta^0-\tilde{\vect \beta}^1}^2\right)-\mathbb{E}\left(\norm{\vect \beta^1-\tilde{\vect \beta}^2}^2\right)\le &2E\left(\norm{\vect \beta^0-\tilde{\vect \beta}^1}\norm{\vect \beta^1-\tilde{\vect \beta}^2-\vect \beta^0+\tilde{\vect \beta}^1}\right)\\
\le& \left(4+4\frac{M}{2}\right)\mathbb{E}(\norm{\vect \beta^0-\tilde{\vect \beta}^1}\norm{\vect \beta^0-\vect \beta^1}).
\end{align*}
By Proposition \ref{pro:expectation} and AM-GM inequality, which for any $b_1,b_2>0$ and any $c>0$, we have
\begin{equation}
\label{equ:AM-GM}
b_1b_2\le \frac{1}{2}\left(cb_1^2+c^{-1}b_2^2\right)
\end{equation}
Therefore, we have
\begin{align*}
\mathbb{E}\left(\norm{\vect \beta^0-\tilde{\vect \beta}^1}\norm{\vect \beta^0-\vect \beta^1}\right)\le & \frac{1}{2}\mathbb{E}\left(\sqrt{n}\norm{\vect \beta^0-\vect \beta^1}^2+\frac{1}{\sqrt{n}}\norm{\vect \beta^0-\tilde{\vect \beta}^1}^2\right)\\
=&\frac{1}{2}\mathbb{E}\left(\frac{1}{\sqrt{n}}\norm{\vect \beta^0-\tilde{\vect \beta}^1}^2+\frac{1}{\sqrt{n}}\norm{\vect \beta^0-\tilde{\vect \beta}^1}^2\right)\;\; \mathrm{(Proposition \ref{pro:expectation})}\\
=&\frac{1}{\sqrt{n}}\mathbb{E}\left(\norm{\vect \beta^0-\tilde{\vect \beta}^1}^2\right)
\end{align*}
Therefore,
\begin{equation*}
\mathbb{E}\left(\norm{\vect \beta^0-\tilde{\vect \beta}^1}^2\right)-\mathbb{E}\left(\norm{\vect \beta^1-\tilde{\vect \beta}^2}^2\right)\le \left(\frac{4}{\sqrt{n}}+\frac{4M}{\sigma\sqrt{n}}\right)\mathbb{E}\left(\norm{\vect \beta^0-\tilde{\vect \beta}^1}^2\right),
\end{equation*}
which implies
\begin{equation*}
\mathbb{E}\left(\norm{\vect \beta^0-\tilde{\vect \beta}^1}^2\right)\le \left(1-\frac{4}{\sqrt{n}}-\frac{4M}{\sigma\sqrt{n}}\right)^{-1}\mathbb{E}\left(\norm{\vect \beta^1-\tilde{\vect \beta}^2}^2\right)\le \rho \mathbb{E}\left(\norm{\vect \beta^1-\tilde{\vect \beta}^2}^2\right),
\end{equation*}
where the last inequality is base on Proposition \ref{pro:rho} and the fact $\theta M\ge 1$.

\textbf{Induction Step.} By the induction hypothesis, we assume
\begin{equation}
\label{equ:rho_t}
\mathbb{E}\left(\norm{\vect \beta^{t-1}-\tilde{\vect \beta}^t}^2\right)\le \rho \mathbb{E}\left(\norm{\vect \beta^{t}-\tilde{\vect \beta}^{t+1}}^2\right)\;\forall t\le l-1.
\end{equation}
To show
\begin{equation*}
\mathbb{E}\left(\norm{\vect \beta^{l-1}-\tilde{\vect \beta}^l}^2\right)\le \rho \mathbb{E}\left(\norm{\vect \beta^{l}-\tilde{\vect \beta}^{l+1}}^2\right),
\end{equation*}
we firstly show that for all $t<j$,
\begin{align}
\label{equ:alpha_t}
&\mathbb{E}\left(\norm{\vect \beta^t-\vect \beta^{t+1}}\norm{\vect \beta^{l-1}-\tilde{\vect \beta}^l}\right)\nonumber\\
\le& \frac{1}{2}\mathbb{E}\left(\sqrt{n}\rho^{(t+1-l)/2}\norm{\vect \beta^t-\vect \beta^{t+1}}^2+\frac{1}{\sqrt{n}}\rho^{(l-1-t)/2}\norm{\vect \beta^{l-1}-\tilde{\vect \beta}^l}^2\right)\;\;\mathrm{(Eq.~\ref{equ:AM-GM})}\nonumber\\
=& \frac{1}{2}\mathbb{E}\left(\sqrt{n}\rho^{(t+1-l)/2}\mathbb{E}\left(\norm{\vect \beta^t-\vect \beta^{t+1}}^2\right)+\frac{1}{\sqrt{n}}\rho^{(l-1-t)/2}\norm{\vect \beta^{l-1}-\tilde{\vect \beta}^l}^2\right)\nonumber\\
=& \frac{1}{2}\mathbb{E}\left(\frac{1}{\sqrt{n}}\rho^{(t+1-l)/2}\norm{\vect \beta^t-\tilde{\vect \beta}^{t+1}}^2+\frac{1}{\sqrt{n}}\rho^{(l-1-t)/2}\norm{\vect \beta^{l-1}-\tilde{\vect \beta}^l}^2\right)\;\; \mathrm{(Proposition~\ref{pro:expectation})}\nonumber\\
\le & \frac{1}{2}\mathbb{E}\left(\frac{1}{\sqrt{n}}\rho^{(t+1-l)/2}\rho^{l-t-1}\norm{\vect \beta^{l-1}-\tilde{\vect \beta}^{l}}^2+\frac{1}{\sqrt{n}}\rho^{(l-1-t)/2}\norm{\vect \beta^{l-1}-\tilde{\vect \beta}^l}^2\right)\;\; \mathrm{(Eq.~\ref{equ:rho_t})}\nonumber\\
\le& \frac{\rho^{(l-1-t)/2}}{\sqrt{n}} \mathbb{E}\left(\norm{\vect \beta^{l-1}-\tilde{\vect \beta}^{l}}^2\right).
\end{align}
Let $\theta=\sum_{t=1}^{\gamma}\rho^{t/2}$. We have
\begin{align*}
&\mathbb{E}\left(\norm{\vect \beta^{l-1}-\tilde{\vect \beta}^l}^2\right)-\mathbb{E}\left(\norm{\vect \beta^{l}-\tilde{\vect \beta}^{l+1}}\right)\\
\le& \mathbb{E}\left(2\norm{\vect \beta^{l-1}-\tilde{\vect \beta}^l}\left(\left(2+2\frac{M}{\sigma}\right)\norm{\vect \beta^{l-1}-\vect \beta^l}+2\frac{M}{\sigma}\norm{\vect \beta^{t-1}-\vect \beta^t}\right)\right)\;\;\mathrm{(Eq.~\ref{equ:alpha_inequality},Eq.~\ref{equ:lemma_alpha_1})}\\
=&\left(4+4\frac{M}{\sigma}\right)\mathbb{E}\left(\norm{\vect \beta^{l-1}-\tilde{\vect \beta}^l}\norm{\vect \beta^{l-1}-\vect \beta^l}\right)+4\frac{M}{\sigma}\sum_{t=l-\gamma-1}^{l-1}\mathbb{E}\left(\norm{\vect \beta^{l-1}-\tilde{\vect \beta}^l}\norm{\vect \beta^{t-1}-\vect \beta^t}\right)\\
\le & \frac{4\sigma+4M}{\sigma\sqrt{n}}\mathbb{E}\left(\norm{\vect \beta^{l-1}-\tilde{\vect \beta}^l}^2\right)+\frac{4M}{\sigma\sqrt{n}}\mathbb{E}\left(\norm{\vect \beta^{l-1}-\tilde{\vect \beta}^l}\right)\sum_{t=l-\gamma-1}^{l-2}\rho^{(l-1-t)/2} \;\;\mathrm{(Eq.~\ref{equ:alpha_t})}\\
\le &\frac{4\sigma+4M}{\sigma\sqrt{n}}\mathbb{E}\left(\norm{\vect \beta^{l-1}-\tilde{\vect \beta}^l}^2\right)+\frac{4M}{\sigma\sqrt{n}}\theta \mathbb{E}\left(\norm{\vect \beta^{l-1}-\tilde{\vect \beta}^l}\right)\\
\le & \left(\frac{4}{\sqrt{n}}+\frac{4M+4M\theta}{\sigma\sqrt{n}}\right)\mathbb{E}\left(\norm{\vect \beta^{l-1}-\tilde{\vect \beta}^l}^2\right)
\end{align*}
which implies that
\begin{equation*}
\mathbb{E}\left(\norm{\vect \beta^{l-1}-\tilde{\vect \beta}^l}^2\right)\le \frac{1}{1-\frac{4}{\sqrt{n}}-\frac{4M+4M\theta}{\sigma\sqrt{n}}}\mathbb{E}\left(\norm{\vect \beta^{l}-\tilde{\vect \beta}^{l+1}}^2\right)\le \rho \mathbb{E}\left(\norm{\vect \beta^{l-1}-\tilde{\vect \beta}^l}^2\right)
\end{equation*}
by Proposition \ref{pro:rho}.
\end{proof}
\subsection{Proof of Lemma \ref{thm:Theta}}
\begin{proof}
We also omit the subscript $_{[k]}$ of the notations in the proof. We can bound the expected distance $\mathbb{E}\left(\norm{\bar{\vect \beta}^{j+1}-\tilde{\vect \beta}^{l+1}}^2\right)$ by the following derivation.
\begin{align}
\label{equ:beta_alpha_tilde}
\mathbb{E}\left(\norm{\bar{\vect \beta}^{l+1}-\tilde{\vect \beta}^{l+1}}^2\right)=&\mathbb{E}\left(\sum_{t=1}^n\left(T_t(\bar{\vect w}^l,\beta^l_t)-T_t(\hat{\vect w}^l,\beta^l_t)\right)^2\right)\nonumber\\
\le & \mathbb{E}\left(\sum_{t=1}^n\left(\frac{\lambda n\left(\bar{\vect w}^l-\hat{\vect w}^l\right)^\top\vect x_t}{\sigma\norm{\vect x_t}^2}\right)^2\right) \;\;\mathrm{(Proposition~\ref{pro:T_w_s})}\nonumber\\
=&\frac{\lambda^2 n^2}{\sigma^2}\mathbb{E}\left(\norm{\bar{\matrx X}\left(\bar{\vect w}^l-\hat{\vect w}^l\right)}^2\right)\nonumber\\
\le & \frac{M^2}{\lambda^2 n^2}\frac{\lambda^2 n^2}{\sigma^2}\mathbb{E}\left(\left(\sum_{t=l-\gamma}^{l-1}\norm{\vect \beta^t-\vect \beta^{t+1}}\right)^2\right) \;\;\mathrm{(Proposition~\ref{pro:M})}\nonumber\\
\le & \frac{M^2}{\sigma^2}\mathbb{E}\left(\gamma\left(\sum_{t=l-\gamma}^{l-1}\norm{\vect \beta^t-\vect \beta^{t+1}}^2\right)\right) \;\;\mathrm{(Cauchy~Schwarz~Inequality)}\nonumber\\
\le & \frac{\gamma M^2}{\sigma^2}\mathbb{E}\left(\gamma\left(\sum_{t=1}^{\gamma}\rho^t\norm{\vect \beta^l-\vect \beta^{l+1}}^2\right)\right) \;\;\mathrm{(Lemma~\ref{thm:lemma_alpha})}\nonumber\\
\le & \frac{\gamma M^2}{\sigma^2n}\left(\sum_{t=1}^{\gamma}\rho^t\right)\mathbb{E}\left(\norm{\vect \beta^l-\tilde{\vect \beta}^{l+1}}^2\right) \;\;\mathrm{(Proposition~\ref{pro:expectation})}\nonumber\\
\le & \frac{\gamma^2 M^2}{\sigma^2n}\rho^\gamma \mathbb{E}\left(\norm{\vect \beta^l-\tilde{\vect \beta}^{l+1}}^2\right)\nonumber\\
\le & \frac{\gamma^2 M^2e^2}{\sigma^2n} \mathbb{E}\left(\norm{\vect \beta^l-\tilde{\vect \beta}^{l+1}}^2\right). \;\;\mathrm{(Proposition~\ref{pro:rho})}\\
\end{align}
Moreover,
\begin{align}
\label{equ:beta_alpha}
\mathbb{E}\left(\norm{\bar{\vect \beta}^l-\vect \beta^{l+1}}^2\right)=&\mathbb{E}\left(\norm{\bar{\vect \beta}^{l+1}-\tilde{\vect \beta}^{l+1}+\tilde{\vect \beta}^{l+1}-\vect \beta^l}^2\right)\nonumber\\
\le &\mathbb{E}\left(2\left(\norm{\bar{\vect \beta}^{l+1}-\tilde{\vect \beta}^{l+1}}^2+\norm{\tilde{\vect \beta}^{l+1}-\vect \beta^l}^2\right)\right) \;\;\mathrm{(Cauchy-Schwarz)}\nonumber\\
\le &2\left(1+\frac{\gamma^2 M^2e^2}{\sigma^2n}\right)\mathbb{E}\left(\norm{\tilde{\vect \beta}^{l+1}-\vect \beta^l}^2\right)
\end{align}
The bound of the increase of local objective function value by
\begin{align*}
&\mathbb{E}\left(D\left(\vect \beta^{l+1}\right)\right)-\mathbb{E}\left(D\left(\vect \beta^{l}\right)\right)\\
=& \mathbb{E}\left(-\left(D\left(\vect \beta^l\right)-D\left(\bar{\vect \beta}^{l+1}\right)\right)\right)-\mathbb{E}\left(\left(D\left(\bar{\vect \beta}^{l+1}\right)-D\left(\vect \beta^{l+1}\right)\right)\right)\\
\ge &\mathbb{E}\left(\frac{\sigma\norm{\vect x_{i(l)}}^2}{2}\norm{\vect \beta^l-\bar{\vect \beta}^{l+1}}^2\right)-\mathbb{E}\left(\frac{L_{max}}{2}\norm{\vect \beta^{l+1}-\bar{\vect \beta}^{l+1}}^2\right) \;\;\mathrm{(Proposition~\ref{pro:convex})} \\
\ge &\frac{R_{min}}{2n}\mathbb{E}\left(\norm{\vect \beta^l-\bar{\vect \beta}^{l+1}}^2\right)-\frac{L_{max}}{2n}\mathbb{E}\left(\norm{\tilde{\vect \beta}^{l+1}-\bar{\vect \beta}^{l+1}}^2\right)\\
\ge & \frac{R_{min}}{2n}\mathbb{E}\left(\norm{\vect \beta^l-\bar{\vect \beta}^{l+1}}^2\right)-\frac{L_{max}}{2n}\frac{\gamma^2 M^2e^2}{\sigma^2n}\mathbb{E}\left(\norm{\tilde{\vect \beta}^{l+1}-\vect \beta^l}^2\right) \;\;\mathrm{(Eq.~\ref{equ:beta_alpha_tilde})}\\
\ge & \frac{R_{min}}{2n}\mathbb{E}\left(\norm{\vect \beta^l-\bar{\vect\beta}^{l+1}}^2\right)-\frac{2L_{max}}{2n}\frac{\gamma^2 M^2e^2}{\sigma^2n}\left(1+\frac{\gamma^2 M^2e^2}{\sigma^2n}\right)\mathbb{E}\left(\norm{\bar{\vect \beta}^{l+1}-\vect \beta^l}^2\right) \;\;\mathrm{(Eq.~\ref{equ:beta_alpha})}\\
\ge &\frac{R_{min}}{2n}\left(1-\frac{2L_{max}}{2n}\left(1+\frac{\gamma^2 M^2e^2}{\sigma^2n}\right)\left(\frac{\gamma^2 M^2e^2}{\sigma^2n}\right)\right)\mathbb{E}\left(\norm{\bar{\vect \beta}^{l+1}-\vect \beta^l}^2\right)\\
\ge &\frac{\kappa R_{min}}{2n}\left(1-\frac{2L_{max}}{2n}\left(1+\frac{\gamma^2 M^2e^2}{\sigma^2n}\right)\left(\frac{\gamma^2 M^2e^2}{\sigma^2n}\right)\right)\mathbb{E}\left(\norm{P_S\left(\vect \beta^l\right)-\vect \beta^l}^2\right) \\
\ge &\frac{\kappa R_{min}}{2nL_{max}}\left(1-\frac{2L_{max}}{2n}\left(1+\frac{\gamma^2 M^2e^2}{\sigma^2n}\right)\left(\frac{\gamma^2 M^2e^2}{\sigma^2n}\right)\right)\mathbb{E}\left(D^*-D\left(\vect \beta^l\right)\right)
\end{align*}
Therefore,
\begin{align*}
D^*-\mathbb{E}\left(D\left(\vect \beta^{l+1}\right)\right)=& D^*-\mathbb{E}\left(D\left(\vect \beta^l\right)\right)-\left(\mathbb{E}\left(D\left(\vect \beta^{l+1}\right)-\mathbb{E}\left(D\left(\vect \beta^l\right)\right)\right)\right)\\
\le &\eta\left(D^*-\mathbb{E}\left(D\left(\vect \beta^l\right)\right)\right)
\end{align*}
\end{proof}
\subsection{Proof of $\Theta$-approximate solution of Lemma \ref{thm:Theta}}
\begin{proof}
	Let us assume that $\beta^{*}_{[k]}$ is the optimal solution of the subproblem \ref{equ:local_opt} denotes as:
	\begin{equation}
		\label{equ:alpha*}
		\beta^{*}_{[k]}=\arg\max\limits_{\beta_{[k]}\in \mathbb{R}^{n_k}} D(\beta_{[k]};\bar{\vect w}).
	\end{equation}
	According to above proof of Lemma \ref{thm:Theta}, the local atomic solver has a linear convergence rate in expectation, that is,
	\begin{equation*}
		D(\beta^{*}_{[k]};\bar{\vect w})-\mathbb{E}\left(D(\beta^{j+1}_{[k]};\bar{\vect w})\right)\le \eta\left(\mathbb{E}\left(D(\beta^{*}_{[k]};\bar{\vect w})-D(\beta^{j}_{[k]};\bar{\vect w})\right)\right)
	\end{equation*}
	It is obvious that $\Theta=\eta^H$. Thus, we can easily get the induction as 
	\begin{align*}
		& D(\beta^{*}_{[k]};\bar{\vect w})-\mathbb{E}\left(D(\beta^{H}_{[k]};\bar{\vect w})\right)\le \eta\left(D(\beta^{*}_{[k]};\bar{\vect w})-\mathbb{E}\left(D(\beta^{H-1}_{[k]};\bar{\vect w})\right)\right) \nonumber \\
		\le &\eta^2\left(\mathbb{E}\left(D(\beta^{*}_{[k]};\bar{\vect w})-D(\beta^{H-2}_{[k]};\bar{\vect w})\right)\right) \le \dots \le \Theta\left(D(\beta^{*}_{[k]};\bar{\vect w})-\mathbb{E}\left(D(\beta^{0}_{[k]};\bar{\vect w})\right)\right).
	\end{align*}
	Notice that $\beta^0_{[k]}$ are the start points of the local atomic solver and $\beta^H_{[k]}$ are the final results of $\beta_{[k]}$ of the local atomic solver. So the following equations hold for the global problem:
	\begin{align*}
		\beta^0_{[k]}&=\alpha_{[k]} \\
		\beta^H_{[k]}-\beta^0_{[k]}&=\vect \delta_{[k]}\\
		\beta^*_{[k]}-\beta^0_{[k]}&=\vect \delta^*_{[k]}
	\end{align*}
	Therefore, we have:
	\begin{equation*}
		\mathbb{E}\left[ Q_{k}^{\sigma}(\gvec{\delta}^*_{[k]}; \vect v, \gvec{\alpha}_{[k]}) - Q_{k}^{\sigma}(\gvec{\delta}_{[k]}; \vect v, \gvec{\alpha}_{[k]}) \right] 	\le \Theta \left[ Q_{k}^{\sigma}(\gvec{\delta}^*_{[k]}; \vect v, \gvec{\alpha}_{[k]}) - Q_{k}^{\sigma}(\bvec{0}; \vect v, \gvec{\alpha}_{[k]}) \right]
	\end{equation*}
	with $\Theta=\eta^H$.
\end{proof}

\section{Proof of convergence of global solution}
\subsection{Proof of Lemma \ref{thm:D_alpha+1}}
\begin{proof}
	Assume that $\mathcal{I}=\bigcup_{k\in\mathcal{P}_S}\mathcal{I}_k$. Then, we have
	\begin{align*}
	D\left(\vect \alpha + \nu\sum_{k\in \mathcal{P}_S}\vect \delta_{[k]}\right)=&-\frac{1}{n}\sum_{i=1}^n\phi_i^*\left(-\alpha_i-\nu\left(\sum_{k\in\mathcal{P}_S}\vect \delta_{[k]}\right)_i\right)\\
	&-\frac{\lambda}{2}\norm{\frac{1}{\lambda n}\matrx X\left(\vect \alpha+\nu\sum_{k\in\mathcal{P}_S}\vect \delta_{[k]}\right)}^2\\
	=&-\frac{1}{n}\sum_{i\notin \mathcal{I}}\phi_i^*(-\alpha_i)-\frac{1}{n}\sum_{k\in\mathcal{P}_S}\left(\sum_{i\in\mathcal{I}_k}\phi_i^*(-(1-\nu)\alpha_i-\nu(\vect \alpha+\vect \delta_{[k]})_i)\right)\\
	&-\frac{\lambda}{2}\left(\norm{\vect w(\vect \alpha)}^2+\frac{2\nu}{\lambda n}\sum_{k\in\mathcal{P}_S}\vect w(\vect \alpha)^\top\matrx X\vect \delta_{[k]}+\left(\frac{\nu}{\lambda n}\right)^2\norm{\sum_{k\in\mathcal{P}_S}\matrx X\vect \delta_{[k]}}^2\right)\\
	\ge &-\frac{1}{n}\sum_{k\in\mathcal{P}_S}\left(\sum_{i\in\mathcal{I}_k}((1-\nu)\phi_i^*(-\alpha_i)+\nu\phi^*_i(-(\vect \alpha+\vect \delta_{[k]})_i))\right)\\
	&-\frac{\lambda}{2}\left(\norm{\hat{\vect w}}^2+\frac{2\nu}{\lambda n}\sum_{k\in\mathcal{P}_S}\hat{\vect w}^\top\matrx X\vect \delta_{[k]}+\left(\frac{\nu}{\lambda n}\right)^2\norm{\sum_{k\in\mathcal{P}_S}\matrx X\vect \delta_{[k]}}^2\right)\\
	&-\frac{\nu}{n}\sum_{k\in \mathcal{P}_S}(\vect w-\hat{\vect w})^\top\matrx X\vect \delta_{[k]}-\frac{\lambda}{2}(\norm{\vect w}^2-\norm{\hat{\vect w}}^2)-\frac{1}{n}\sum_{i\notin \mathcal{I}}\phi_i^*(-\alpha_i)\\
	=&\underbrace{-\frac{1}{n}\sum_{k\in\mathcal{P}_S}\left(\sum_{i\in\mathcal{I}_k}(1-\nu)\phi_i^*(-\alpha_i)\right)-(1-\nu)\frac{\lambda}{2}\norm{\vect w(\hat{\vect \alpha})}^2}_{(1-\nu)D(\hat{\vect \alpha})}\\
	&+\nu\sum_{k\in\mathcal{P}_k}\left(-\frac{1}{n}\sum_{i\in \mathcal{I}_k}\phi^*_i(-(\vect \alpha+\vect \delta_{[k]})_i)-\frac{1}{K}\frac{\lambda}{2}\norm{\vect w(\hat{\vect \alpha})}^2\right.\\
	&\left.-\frac{1}{n}\vect w(\hat{\vect \alpha})^\top\matrx X\vect \delta_{[k]}-\frac{\lambda}{2}\sigma\norm{\frac{1}{\lambda n}\matrx X \vect \delta_{[k]}}^2\right)\\
	&-\frac{\nu}{n}\sum_{k\in \mathcal{P}_S}(\vect w-\hat{\vect w})^\top\matrx X\vect \delta_{[k]}-\frac{\lambda}{2}(\norm{\vect w}^2-\norm{\hat{\vect w}}^2)-\frac{1}{n}\sum_{i\notin \mathcal{I}}\phi_i^*(-\alpha_i)\\
	=& (1-\nu)D(\hat{\vect \alpha})+\nu\sum_{k\in \mathcal{P}_S}Q^\sigma_k(\vect \delta_{[k]};\vect \alpha_{[k]},\hat{\vect w})\\
	&-\frac{\nu}{n}\sum_{k\in \mathcal{P}_S}(\vect w-\hat{\vect w})^\top\matrx X\vect \delta_{[k]}-\frac{\lambda}{2}(\norm{\vect w}^2-\norm{\hat{\vect w}}^2)-\frac{1}{n}\sum_{i\notin \mathcal{I}}\phi_i^*(-\alpha_i).
	\end{align*}	
\end{proof}

\subsection{Proof of Lemma~\ref{lem:global_convergence}}
\begin{proof}
	For sake of notation, we will write $\vect \alpha$ instead of $\vect \alpha^t$, $\vect w$ instead of $\vect w(\vect \alpha^t)$, $\hat{\vect w}$ instead of $\vect w(\hat{\vect \alpha})$, and $\vect \delta$ instead of $\vect \delta^t$. 
	
	Now, the expected change of the dual objective is
	\begin{align*}
	\mathbb{E}[D(\vect \alpha^t)-D(\vect \alpha^{(t+1)})]=&\mathbb{E}[D(\vect \alpha^t)-D(\hat{\vect \alpha})+D(\hat{\vect \alpha})-D(\vect \alpha^{(t+1)})]\\
	=&\mathbb{E}[D(\vect \alpha^t)-D(\hat{\vect \alpha})]+\mathbb{E}[D(\hat{\vect \alpha})-D(\vect \alpha^{(t+1)})]
	\end{align*}
	Thus, it is a summation of two parts. Let us estimate both parts as following,
	\begin{align*}
	\mathbb{E}[D(\vect \alpha^t)-D(\hat{\vect \alpha})]=&\mathbb{E}\left[-\frac{1}{n}\sum_{i\notin \mathcal{I}}\phi_i^*(-\alpha_i)-\frac{1}{n}\sum_{k\in\mathcal{P}_S}\left(\sum_{i\in\mathcal{I}_k}\phi^*_i(-\alpha_i)\right)-\frac{\lambda}{2}\norm{\vect w}^2+\frac{1}{n}\sum_{k\in\mathcal{P}_S}\left(\sum_{i\in\mathcal{I}_k}\phi^*_i(-\alpha_i)\right)+\frac{\lambda}{2}\norm{\hat{\vect w}}^2\right]\\
	=&\mathbb{E}\left[-\frac{\lambda}{2}\left(\norm{\vect w}^2-\norm{\hat{\vect w}}^2\right)-\frac{1}{n}\sum_{i\notin \mathcal{I}}\phi_i^*(-\alpha_i)\right]
	\end{align*}
	
	\begin{align*}
	\mathbb{E}[D(\hat{\vect \alpha})-D(\vect \alpha^{(t+1)})]=&\mathbb{E}\left[D(\hat{\vect \alpha})-D(\vect \alpha+\nu\sum_{k\in\mathcal{P}_S}\vect \delta_{[k]})\right]\\
	\le & \mathbb{E}\left[D(\hat{\vect \alpha})-(1-\nu)D(\hat{\vect \alpha})-\nu\sum_{k\in\mathcal{P}_S}Q^\sigma_k(\vect \delta_{[k]};\vect \alpha_{[k]},\hat{\vect w})\right]\\
	&+\mathbb{E}\left[\frac{\nu}{n}\sum_{k\in \mathcal{P}_S}(\vect w-\hat{\vect w})^\top\matrx X\vect \delta_{[k]}+\frac{\lambda}{2}(\norm{\vect w}^2-\norm{\hat{\vect w}}^2)+\frac{1}{n}\sum_{i\notin \mathcal{I}}\phi_i^*(-\alpha_i)\right]\;\;\mathrm{(Lemma~\ref{thm:D_alpha+1})}
	\end{align*}
	
	Therefore,
	\begin{align*}
	\mathbb{E}[D(\vect \alpha^t)-D(\vect \alpha^{(t+1)})]\le&\mathbb{E}\left[-\frac{\lambda}{2}\left(\norm{\vect w}^2-\norm{\hat{\vect w}}^2\right)-\frac{1}{n}\sum_{i\notin \mathcal{I}}\phi_i^*(-\alpha_i)\right]\\
	&+\mathbb{E}\left[D(\hat{\vect \alpha})-(1-\nu)D(\hat{\vect \alpha})-\nu\sum_{k\in\mathcal{P}_S}Q^\sigma_k(\vect \delta_{[k]};\vect \alpha_{[k]},\hat{\vect w})\right]\\
	&+\mathbb{E}\left[\frac{\nu}{n}\sum_{k\in \mathcal{P}_S}(\vect w-\hat{\vect w})^\top\matrx X\vect \delta_{[k]}+\frac{\lambda}{2}(\norm{\vect w}^2-\norm{\hat{\vect w}}^2)+\frac{1}{n}\sum_{i\notin \mathcal{I}}\phi_i^*(-\alpha_i)\right]\\
	=&\nu\mathbb{E}\left[D(\hat{\vect \alpha})-\sum_{k\in\mathcal{P}_S}Q^\sigma_k(\vect \delta^*_{[k]};\vect \alpha_{[k]},\hat{\vect w})\right.\\
	&\left.+\sum_{k\in\mathcal{P}_S}Q^\sigma_k(\vect \delta^*_{[k]};\vect \alpha_{[k]},\hat{\vect w})-\sum_{k\in\mathcal{P}_S}Q^\sigma_k(\vect \delta_{[k]};\vect \alpha_{[k]},\hat{\vect w})\right]\\
	&+\mathbb{E}\left[\frac{\nu}{n}\sum_{k\in \mathcal{P}_S}(\vect w-\hat{\vect w})^\top\matrx X\vect \delta_{[k]}\right]\\
	\le & \nu\bigg(D(\hat{\vect \alpha})-\sum_{k\in\mathcal{P}_S}Q^\sigma_k(\vect \delta^*_{[k]};\vect \alpha_{[k]},\hat{\vect w})\\
	&+\Theta\Big(\sum_{k\in\mathcal{P}_S}Q^\sigma_k(\vect \delta^*_{[k]};\vect \alpha_{[k]},\hat{\vect w})-\underbrace{\sum_{k\in\mathcal{P}_S}Q^\sigma_k(\vect 0;\vect \alpha_{[k]},\hat{\vect w})}_{D(\hat{\vect \alpha})}\Big)\bigg)\;\;\mathrm{(Lemma~\ref{thm:convergence_Theta})}\\
	&+\mathbb{E}\left[\frac{\nu}{n}\sum_{k\in \mathcal{P}_S}(\vect w-\hat{\vect w})^\top\matrx X\vect \delta_{[k]}\right]\\
	=&\nu(1-\Theta)\left(D(\hat{\vect \alpha})-\sum_{k\in\mathcal{P}_S}Q^\sigma_k(\vect \delta^*_{[k]};\vect \alpha_{[k]},\hat{\vect w})\right)+\mathbb{E}\left[\frac{\nu}{n}\sum_{k\in \mathcal{P}_S}(\vect w-\hat{\vect w})^\top\matrx X\vect \delta_{[k]}\right]\\
	\le &\nu(1-\Theta)\left(D(\hat{\vect \alpha})-\sum_{k\in\mathcal{P}_S}Q^\sigma_k(\vect \delta^*_{[k]};\vect \alpha_{[k]},\hat{\vect w})\right)\\
	&+\frac{\nu}{2n}\underbrace{\left(\mathbb{E}\left[\norm{\vect w-\hat{\vect w}}^2\right]+\mathbb{E}\left[\norm{\sum_{k\in \mathcal{P}_S}\matrx X\vect \delta_{[k]}}^2\right]\right)}_{A}
	\end{align*}
	Note that $\vect w=\hat{\vect w}+\frac{1}{\lambda n}\sum_{j=t-\Gamma}^{t-1}\matrx X\vect \delta^j$. Now, let us bound the term $A$. We have
	\begin{align*}
	A=&\mathbb{E}\left[\frac{1}{\lambda n}\norm{\sum_{j=t-\Gamma}^{t-1}\matrx X\vect \delta^j}^2\right]+\mathbb{E}\left[\norm{\sum_{k\in \mathcal{P}_S}\matrx X\vect \delta_{[k]}}^2\right]\\
	\le & \mathbb{E}\left[\frac{\Gamma}{\lambda n}\sum_{j=t-\Gamma}^{t-1}\norm{\matrx X\vect \delta^j}^2\right]+\mathbb{E}\left[S\sum_{k\in \mathcal{P}_S}\norm{\matrx X\vect \delta_{[k]}}^2\right]\;\;\mathrm{(Cauchy~Schwarz~Inequality)}\\
	\le & \mathbb{E}\left[\frac{\Gamma M}{\lambda n}\sum_{j=t-\Gamma}^{t-1}\norm{\vect \delta^j}^2\right]+\mathbb{E}\left[SM\sum_{k\in \mathcal{P}_S}\norm{\vect \delta_{[k]}}^2\right]\;\;\mathrm{(Proposition~\ref{pro:M})}\\
	\le & \mathbb{E}\left[\frac{\Gamma M}{\lambda n}\left(\sum_{j=t-\Gamma}^{t-1}\varrho^j\right)\norm{\vect \delta^{t-1}}^2\right]+\mathbb{E}\left[SM\sum_{k\in \mathcal{P}_S}\norm{\vect \delta_{[k]}}^2\right]\;\;\mathrm{(By~(\ref{equ:varrho_bound}))}\\
	\le & \frac{\Gamma ML_{max}}{2\lambda n}\sum_{j=t-\Gamma}^{t-1}\varrho^j\left(D(\hat{\vect \alpha})-\sum_{k\in \mathcal{P}_S}Q^\sigma_k(\vect \delta_{[k]};\vect \alpha_{[k]},\hat{\vect w})\right)\\
	&+\frac{SML_{max}}{2}\sum_{k\in \mathcal{P}_S}\left(D(\hat{\vect \alpha})-Q^\sigma_k(\vect \delta_{[k]};\vect \alpha_{[k]},\hat{\vect w})\right)\;\;\mathrm{(Proposition~\ref{pro:convex})}\\
	\end{align*}
	Here $D(\hat{\vect \alpha})=Q^\sigma_k(\vect 0;\vect \alpha_{[k]},\hat{\vect w})$. Thus, Eq. \ref{eq:subopt} can be rewritten as,
	\begin{align}
	\label{equ:Theta2}
	\mathbb{E}\left[Q^\sigma_k(\vect \delta^*_{[k]};\vect \alpha_{[k]},\hat{\vect w})-Q^\sigma_k(\vect \delta_{[k]};\vect \alpha_{[k]},\hat{\vect w})\right]\le& \Theta\left(Q^\sigma_k(\vect \delta^*_{[k]};\vect \alpha_{[k]},\hat{\vect w})-D(\hat{\vect \alpha})\right)+D(\hat{\vect \alpha})-D(\hat{\vect \alpha})\nonumber\\
	D(\hat{\vect \alpha})-Q^\sigma_k(\vect \delta_{[k]};\vect \alpha_{[k]},\hat{\vect w})\le&(1-\Theta)D(\hat{\vect \alpha})-(1-\Theta)Q^\sigma_k(\vect \delta^*_{[k]};\vect \alpha_{[k]},\hat{\vect w})\nonumber\\
	D(\hat{\vect \alpha})-Q^\sigma_k(\vect \delta_{[k]};\vect \alpha_{[k]},\hat{\vect w})\le&-(1-\Theta)\left(Q^\sigma_k(\vect \delta^*_{[k]};\vect \alpha_{[k]},\hat{\vect w})-D(\hat{\vect \alpha})\right)
	\end{align}
	Then, $A$ can be bounded as,
	\begin{align*}
	A\le & -\frac{\Gamma^2 ML_{max}}{2\lambda n}\varrho^\Gamma(1-\Theta)\sum_{k\in \mathcal{P}_S}\left(D(\hat{\vect \alpha})-Q^\sigma_k(\vect \delta^*_{[k]};\vect \alpha_{[k]},\hat{\vect w})\right)\\
	&-\frac{SML_{max}}{2}(1-\Theta)\sum_{k\in \mathcal{P}_S}\left(D(\hat{\vect \alpha})-Q^\sigma_k(\vect \delta^*_{[k]};\vect \alpha_{[k]},\hat{\vect w})\right)\;\;\mathrm{(By~(\ref{equ:Theta2}))}\\
	\le & -\frac{\Gamma^2 eML_{max}}{2\lambda n}(1-\Theta)\sum_{k\in \mathcal{P}_S}\left(D(\hat{\vect \alpha})-Q^\sigma_k(\vect \delta^*_{[k]};\vect \alpha_{[k]},\hat{\vect w})\right)\\
	&-\frac{SML_{max}}{2}(1-\Theta)\sum_{k\in \mathcal{P}_S}\left(D(\hat{\vect \alpha})-Q^\sigma_k(\vect \delta^*_{[k]};\vect \alpha_{[k]},\hat{\vect w})\right)\;\;\mathrm{(Assumption~(\ref{as:bounded_rho}))}
	\end{align*}
	By substituting $A$, we have
	\begin{align*}
	\mathbb{E}[D(\vect \alpha^t)-D(\vect \alpha^{(t+1)})]\le&\nu\left(1-\frac{\Gamma^2 eML_{max}}{4\lambda n^2}-\frac{SML_{max}}{4n}\right)(1-\Theta)\sum_{k\in \mathcal{P}_S}\left(D(\hat{\vect \alpha})-Q^\sigma_k(\vect \delta^*_{[k]})\right)\\
	\mathbb{E}[D(\vect \alpha^t)-D(\vect \alpha^{(t+1)})]\le&\Psi(1-\Theta)\sum_{k\in \mathcal{P}_S}\left(D(\hat{\vect \alpha})-Q^\sigma_k(\vect \delta^*_{[k]})\right)
	\end{align*}
	
	Using the Eq.~C in the proof of Lemma 5 in \cite{ma2015adding}, we can show that
	\begin{align*}
	\mathbb{E}[D(\vect \alpha^t)-D(\vect \alpha^{(t+1)})]\le&\Psi(1-\Theta)\left(-sG(\hat{\vect \alpha})-\frac{1}{2\mu}(1-s)s\frac{1}{n}\norm{\hat{\vect u}-\hat{\vect \alpha}}^2+\frac{\sigma}{2\lambda}\left(\frac{s}{n}\right)^2\sum_{k\in\mathcal{P}_S}\norm{\matrx X(\hat{\vect u}-\hat{\vect \alpha})_{[k]}}\right)\\
	= &\Psi(1-\Theta)\left(-sG(\hat{\vect \alpha})+\frac{\sigma}{2\lambda}\left(\frac{s}{n}\right)^2\hat{R}\right)
	\end{align*}
	
\end{proof}

\end{document}